\documentclass[letterpaper]{emulateapj}
\usepackage{bigdelim}
\usepackage{float} 
\usepackage{amsmath}
\usepackage{epsfig,floatflt}
\usepackage{natbib}
\usepackage{subfigure}
\usepackage{amssymb}
\usepackage{color}
\usepackage{multirow}
\usepackage{amsmath}
\usepackage{color}
\usepackage{url}
\usepackage{subfigure}
\usepackage{float}
\usepackage{hhline}

\DeclareMathSymbol{\la}{3}{AMSa}{46}
\DeclareMathSymbol{\ga}{3}{AMSa}{38}

\newcommand{\LCDM}{$\Lambda$CDM }
\newcommand{\be}{\begin{equation}} \newcommand{\ee}{\end{equation}}
\newcommand{\ba}{\begin{eqnarray}} \newcommand{\ea}{\end{eqnarray}}
\newcommand{\brr}{\begin{array}} \newcommand{\err}{\end{array}}
\newcommand{\kms}{km~s$^{-1}$}
\newcommand{\kmsMpc}{km~s$^{-1}$~Mpc$^{-1}$}
\newcommand{\Mpch}{$h^{-1}$~Mpc}

\newcommand{\Msun}{\text{M}_\odot}

\begin{document}

\title{Cosmic flow from 2MASS redshift survey: The origin of CMB dipole and implications for \LCDM{} cosmology}

\author{ Guilhem Lavaux\altaffilmark{1,2}, 
R. Brent Tully\altaffilmark{3},
Roya Mohayaee\altaffilmark{1}, 
St\'ephane Colombi\altaffilmark{1} 
} \altaffiltext{1}{Institut d'Astrophysique de Paris, UMR 7095 CNRS/Universit\'e Pierre et Marie Curie, 98bis bd Arago,
France}
\altaffiltext{2}{University of Illinois, Department of Physics, 1110 W. Green St, Urbana, IL, USA}
\altaffiltext{3}{Institute for Astronomy, Univ. of Hawaii, Honolulu, USA}
\begin{abstract}

We generate the peculiar velocity field for the 2MASS Redshift Survey
(2MRS) catalog \citep{HuchraIAU05} using an orbit-reconstruction algorithm.  The
reconstructed velocities of individual objects in 2MRS are
well-correlated with the peculiar velocities obtained from
high-precision observed distances within 3,000 \kms. We estimate the
mean matter density to be $\Omega_\text{m} = 0.31 \pm 0.05$ by
comparing observed to reconstructed velocities in this volume.  The
reconstructed motion of the Local Group in the rest frame established
by distances within 3,000 \kms{} agrees with the observed motion and is
generated by fluctuations within this volume, in agreement with
observations.  Having tested our method against observed distances, we
reconstruct the velocity field of 2MRS in successively larger radii,
to study the problem of convergence towards the CMB dipole. We find
that less than half of the amplitude of the CMB dipole is generated
within a volume enclosing the Hydra-Centaurus-Norma supercluster at
around 40\Mpch. Although most of the amplitude of the CMB dipole
seems to be recovered by 120\Mpch, the direction does not agree and
hence we observe no convergence up to
this scale. Due to dominant superclusters such as
Shapley or Horologium-Reticulum in the southern hemisphere at scales above
120\Mpch{}, one might need to go well beyond 200\Mpch{} to
fully recover the dipole vector. 

We develop a statistical model which allows us to estimate
cosmological parameters from the reconstructed growth of convergence
of the velocity of the Local Group towards the CMB dipole motion. 
For scales up to 60\Mpch, assuming a Local Group velocity of 627~\kms, we
estimate $\Omega_\text{m} h^2 = 0.11 \pm 0.06$ and $\sigma_8=0.9 \pm 0.4$, in agreement with WMAP5 measurements at the 1-$\sigma$ level. However, for scales up to 100\Mpch, we obtain
$\Omega_\text{m} h^2 = 0.08\pm 0.03$ and $\sigma_8=1.0 \pm 0.4$, which agrees at the 1 to 2-$\sigma$ level with WMAP5 results.
\end{abstract}


\maketitle

\newcommand{\mycite}[1]{\citeauthor{#1} \citeyear{#1}}

\section{Introduction}
\label{sec:introduction}

The Cosmic Microwave Background (CMB) dipole indicates that our Local
Group (LG) moves with a velocity of $627\pm 22$~\kms{} towards the
direction $l=276^\circ$, $b=30^\circ$ in galactic
coordinates. The first attempt at making a comparison
  between the dipole induced by the gravitational influence of
  structures in our Local Universe and the observed CMB dipole was
  made by \cite{Yahil80} using the revised Shapley-Ames catalog and by
  \cite{Davis82} with the CfA catalog. This initial attempt was followed
  by a great number of works \citep{Yahil86,LB89,RR90,Strauss92,Hudson93}. 
  More recently, this comparison
  has been made with the extended source catalog of the 2 Micron
  All-Sky Survey by \cite{Maller03}. These studies all agree within
  10-30$^\circ$ with each other and with the observed direction of the
  CMB dipole. Nowadays, it is considered unlikely that the
Hydra-Centaurus-Norma supercluster at around 40\Mpch{} could be solely
responsible for the dipole. A number of authors suggests that one has
to go at least as far as the Shapley concentration at about 150\Mpch{} in order
to fully recover the dipole motion (\mycite{Kocevski2006}, \mycite{Plionis98},
\mycite{BP96}, \mycite{Strauss92} for gravity induced dipole, \mycite{Scaramella89}, \mycite{Hoffman01} for POTENT analysis of the observed tidal field).
However, due mainly to sparseness of data at very large distances, 
there is still no consensus on the depth of the
convergence for the CMB dipole, as some authors argued for a quicker
convergence (\mycite{Erdogdu2005}, \mycite{Erdogdu2006}, \mycite{LB89},
  \mycite{Lahav87}, \mycite{Lahav88}, \mycite{Strauss92}, \mycite{Yahil86} and \mycite{Lahav89}).
This problem is even more complicated by the fact that we do not know
with a high precision the value of the linear bias for each of these
surveys. This uncertainty gives some freedom on the interpretation of
the apparent convergence of the velocity of the Local Group.  For
example, \cite{RR00} argued that we reached a convergence assuming a
low value for the biasing whereas \cite{Hudson94} argued against
with a higher value for the biasing.

The other approach of studying the convergence of the
velocity of the Local Group is to use surveys of peculiar velocities
of galaxies. \cite{Watkins08} showed that all the recent surveys
give consistently a large bulk flow on a 100\Mpch{} scale. Only
the data of \cite{LP1994} still remains inconsistent.
This seems to indicate that the gravity induced dipole should not have 
converged  by 100\Mpch{}. 

The common approach to recovery of the CMB dipole has been to
use linear theory to reconstruct the velocity field from
redshift surveys (see {\it e.g.} \citeauthor{Erdogdu2005}
\citeyear{Erdogdu2005},\citeyear{Erdogdu2006};
\citeauthor{Kocevski2006} \citeyear{Kocevski2006}). Methods based on
linear theory can suffer from inadequacy in dealing with large
fluctuations such as the Hydra-Centaurus-Norma supercluster, the Perseus-Pisces supercluster and the Virgo cluster. These problems may be enhanced by the coupling with redshift space distortions.
Contrary to \cite{Erdogdu2006}, we also test here
our reconstructed velocity field against distance measurements.

Here, we apply the Lagrangian method, Monge-Amp\`ere-Kantorovitch (MAK), of peculiar-velocity
reconstruction to the 2-Micron All-Sky Redshift Survey (2MRS) catalog \citep{Huchra00,HuchraIAU05,Erdogdu2005,Erdogdu2006} and produce a 3-dimensional map of
the velocity field and study the Local Group velocity convergence.
Compared to \cite{Erdogdu2006}, we account better for non-linear effects
in peculiar velocities, though at the price of having to correct for redshift distortion. The method has been adapted to work directly with
redshifts and allows us to go well beyond the linear theory into the
non-linear regime. The method has been tested previously against
simulations and mock catalogs and has been shown to reconstruct
reliable peculiar velocities on scales above 4-5 Mpc (see
\citeauthor{lavaux08} \citeyear{lavaux08} and references therein). In
addition, using this method we can directly generate the 3-components
of the peculiar velocities and hence overcome projection effects.

The reconstructed velocities of the 2MRS galaxies are tested against
observed peculiar velocities obtained from high-precision distance measurements
within a 3,000~\kms{} (3k) radius \citep{TullyVoid07}.  The
reconstructed and observed velocities are well-correlated with small
dispersion. We make an independent estimate of $\Omega_\text{m}$ using
this comparison which is in excellent agreement with the results
obtained using WMAP5 \citep{WMAP5_LCDM}. The second test consists of
comparing the reconstructed velocity of the Local Group in the rest
frame of the 3,000~\kms{} sample to the observed velocity. Our results show that
the Local Group motion in the rest frame of 3,000~\kms{} is mostly generated
within this volume, as indicated by the observations
\citep{TullyVoid07}.

Having tested our method, we study the origin of the CMB dipole.  We
reconstruct the velocity field of 2MRS in successively larger radii in
order to determine whether there is any trend of convergence towards
the CMB dipole.  Contrary to \cite{Erdogdu2006} but more in agreement
with \cite{Pike05}, we show that the depth of the convergence in 2MRS
lies beyond 120\Mpch.  Due to severe incompleteness of the 2MRS
catalog beyond 120\Mpch, we can only put a lower-limit on the value of
the convergence depth.  Our method allows us to determine the rate at
which the dipole is approached: less than 50\% of the dipole amplitude
is achieved at around the Hydra-Centaurus-Norma distance and at $\sim
120$\Mpch{} we recover about 87\% of the amplitude of the CMB dipole
but with no evidence for convergence in direction. We then compare the
rate of convergence that we measure to theoretical predictions given
by \LCDM{} and the linear theory. It appears that our results are in
agreement with \LCDM{} when we consider galaxies at distances lower
than 60\Mpch. Our results deviate 1$\sigma$ to 2$\sigma$ from
expectations of the cosmology of WMAP5 for reconstruction of larger
radii. A problem relative to the bulk flow on a 100\Mpch{} scale has also
been noted recently by \cite{Watkins08} and on a 300\Mpch{} scale by \cite{Kashlinsky08}. \cite{Watkins08} show that, though the bulk flow found
by \cite{LP1994} is not in agreement with any other peculiar velocity survey
so far, there does exist an excessively large bulk flow, whose amplitude
and direction is compatible with our findings.
In this analysis, the Local Group velocity was taken as a constraint for the
conditional probability density function. We have made a parallel analysis 
by removing this constraint and studying the joint probability distribution function instead of the conditional probability distribution function. We have tested that weighing our likelihood analysis by the probability of occurrence of the
velocity of the Local Group does not change our results.

This paper is organized as follows.  In Section \ref{sec:catalogs}, we
introduce the 2MRS catalog and the 3,000~\kms{} distance catalog and we describe
the catalog that we make by combining these two. In Section
\ref{sec:techniques}, we describe our method of peculiar velocity
reconstruction. In Section~\ref{sec:result1}, we constrain the density parameter $\Omega_\text{m}$ by comparing the reconstructed velocity field of 2MRS with the observed velocities obtained from measured distances within the
3,000~\kms{} volume. In
Section~\ref{sec:result2}, we study the reconstructed evolution of the
velocity of the Local Group towards the CMB dipole. We then discuss,
in Section~\ref{sec:compat_wmap5}, the compatibility of this growth
with the cosmological parameters as given by WMAP5. We develop a
Bayesian analysis in Section~\ref{sec:param_estimate} which allows us
to make an estimate of cosmological parameters using our measurement
on the velocity of the Local Group in different rest frames.  In
Section \ref{sec:conclusion} we conclude.

\section{The combined catalog: 2MASS redshift catalog (2MRS) plus catalog of distances within 3,000~\kms{} (3k distance catalog)}
\label{sec:catalogs}

\begin{figure}
 \begin{center}
   \includegraphics*[width=\hsize]{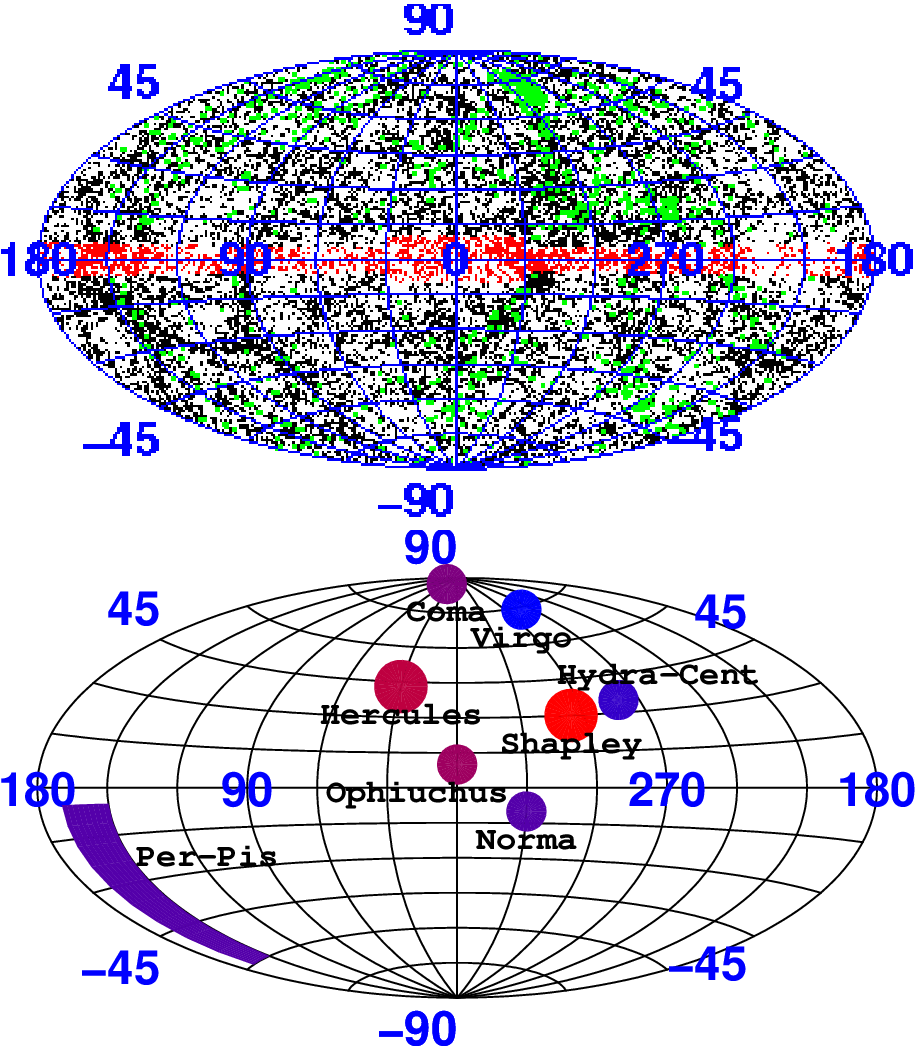}
 \end{center}  
 \caption{\label{fig:2MRS_distance} {\it Combined 2MRS redshift
     catalog and 3k catalog of distances}: Top panel: Large solid
   green circles are the galaxies for which a distance measure is
   available. The zone of avoidance has also been filled and is shown
   by the red belt running across the catalog. Bottom panel: angular
   positions of a few dominant superclusters in 2MRS catalog with
   more distant objects shown by redder circles. We represented the
   position of the Ophiuchus supercluster which has not been included
   in our galaxy sample.}
\end{figure}

The 2MRS \citep{Huchra00,HuchraIAU05} is the most uniformly sampled all-sky redshift survey to 
date and its selection in the near infra-red 
reduces the impact of the zone of avoidance.
The catalog is based on galaxy selection in the infra-red from 
the Two Micron All-Sky Survey (2MASS, \mycite{2MASS}). 
The K=11.25 mag 2MRS is now complete and contains about 23,200 galaxies.
2MRS provides good coverage to a mean distance of 60\Mpch{} and
becomes extremely sparse afterward.

We fill the zone of avoidance, which mainly extends over $|b| <
5-10^\circ$, by mirroring galaxies from the unmasked region (first mentioned
in \citeauthor{LB89} \citeyear{LB89}, see also
\citeauthor{Shay95} \citeyear{Shay95} and \citeauthor{lavaux08}
\citeyear{lavaux08} for details). The mask has been chosen to cover
the part of the sky within $10^\circ$ of the galactic plane for
galactic longitudes $l < 30^\circ$ and $l > 330^\circ$, and within
$5^\circ$ of it for other galactic longitudes. That way, we avoid the
deeply obscured region behind the galactic bulge, at the expense of
losing, {\it e.g.}, the Ophiuchus supercluster. The fake galaxies are
represented by red points in the top panel of
Fig.~\ref{fig:2MRS_distance}. The kinematic fingers of god
\citep{fogJackson}, associated with clusters, are compressed using the
algorithm of \cite{HuchraGeller82}. We used a fiducial velocity
$V_F=1000$~\kms, $V_0 = 350$~\kms, $D_0 = 0.41\,h^{-1}\text{Mpc}$,
which corresponds to detecting galaxy count overdensities of $\delta n
/ \bar{n} \simeq 80$ \citep{Crook07,Crook07err} and provide robust statistical properties of
groups as explained by \cite{Ramella97}. We assume a constant $M/L$
ratio as it has been shown that the masses of big groups of galaxies
only moderately depend on luminosities \citep{ML_2MASS}. Moreover, we
have checked using the virial theorem on big groups of galaxies
(typically $M \ga 10^{12}$~M$_\odot$) that this is the best assumption
we can make given the present data. 
Thus, for the calculation of the velocity of the Local Group,
our method looks both like a flux-weighted computation \citep{LB89}, because we depend on luminosities of galaxies to infer the mass, and
number-weighed \citep{Yahil91}, because we will be dependent on a correction of  redshift distortions.
Incompleteness is taken into account
by using the Schechter luminosity function given in \cite{Crook07} and
\cite{HuchraIAU05} for 2MRS. The specific values of the parameters
are: $\Phi_* = 1.06\, 10^{-2} h^3\text{Mpc}^{-3}$~, $M_* = -24.2 + 5
\log_{10} h_{70}$. We have verified that, as far as galaxies of the
2MRS are concerned, there does not seem to be any systematic effect in
the relation between the $K_{20}$ magnitude and the intrinsic
luminosity.

We chose to execute the reconstruction itself in the CMB
  rest frame. Though it means that redshift distortions may be
  stronger near the observer, this ensures that better boundary
  conditions are enforced on the outer edge of the reconstruction
  volume. Using the Local Group
  rest frame for large reconstruction volumes may lead to the
  equivalent of the so-called {\it rocket effect} \citep{Kaiser87}, which in the context of
  MAK still requires to be properly modeled. Being in the CMB
  rest frame, this effect should be negligible.

The second observational component is an extended catalog of galaxy 
distances.  Information from four techniques has been
integrated: the Cepheid variable \citep{cepheid}, Tip of the Red Giant 
Branch \citep{tip_giant1,tip_giant2}, Surface Brightness Fluctuation 
\citep{sbf1,sbf2}, and 
Luminosity--Linewidth \citep{tully_fisher,Tully00} methods.
In all, there are 1791 galaxies with
distance measures within 3,000~km~s$^{-1}$ (whence the name ``3k catalog''); over 600 of
these are derived by at least one of the first three `high quality'
techniques.  
The 3k catalog of distances has been described in detail by \cite{TullyVoid07}.

The combined 2MRS catalog and 3k distance catalog are 
shown in Figure~\ref{fig:2MRS_distance}.
In the present study, distances are averaged over groups
because our method cannot meaningfully recover the velocities 
on sub-group scales.
The present mixed catalog is composed of 24,819 galaxies, among which 
1,126 has been assigned a distance. Distances are assigned to 109 
groups of galaxies out of 695 groups.  There are 
617 galaxies with measured distances but which have not been grouped.

\begin{figure}
 \begin{center}
   \includegraphics[width=\hsize]{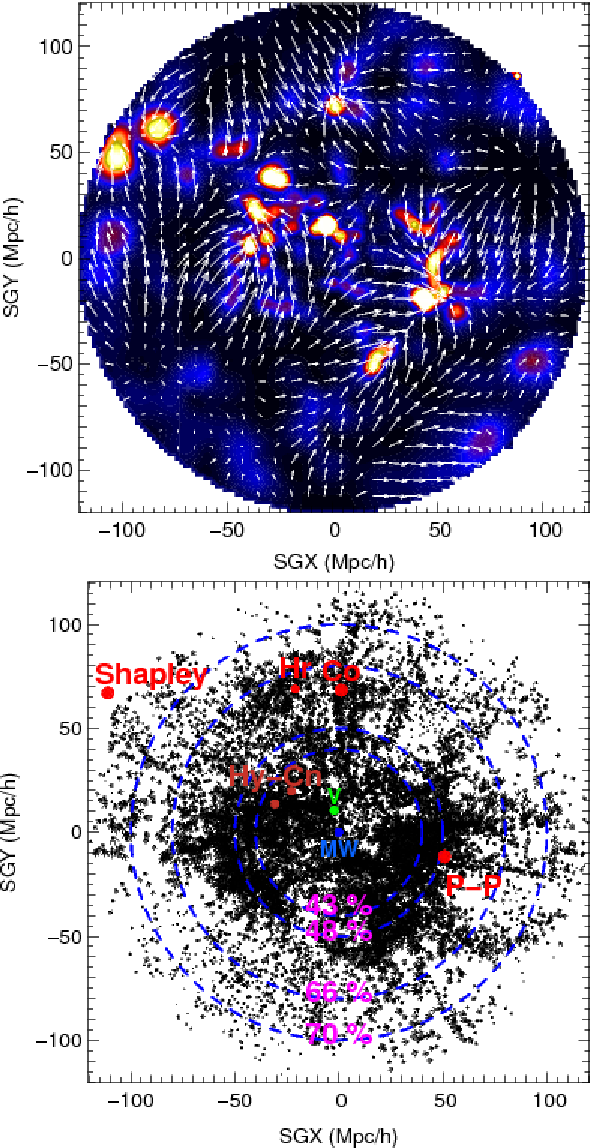}
 \end{center}
 \caption{
{\it 2MRS peculiar velocity field}: Top panel shows a thin slice of
the 2MRS peculiar velocity field adaptively smooth on a grid with
$128^3$ sites sampling a cube with an edge of 240~\Mpch{}. The
velocity field is then subsampled 4 times before being shown. The
underlying density field has been computed by putting objects at their
redshift position.  Bottom panel: Peculiar velocities of individual 2MRS
galaxies in a 40~\Mpch{} slice, centered on the Supergalactic plane
SGZ=0~\kms.  2MRS becomes severely incomplete after 120\Mpch. Hy-Cn
stands for the Hydra-Centaurus supercluster, Hr for Hercules, Co for
Coma, P-P for Perseus-Pisces.
\label{fig:2mrs_velfield}}
\end{figure}

\section{Technique: MAK reconstruction of peculiar velocity field}
\label{sec:techniques}

The technique that will be implemented here
is the Monge-Amp\`ere-Kantorovich ({\small MAK}) 
reconstruction.
This method provides a recipe for reconstructing galaxy orbits
that is unique to the degree that orbits can be described as following
straight lines under suitable coordinate transformations 
\citep[][ and references therein]{lavaux08}. 
In essence, it tries to find the unique displacement
field that does not produce shell crossings and that map an assumed homogeneous
initial density field to the presently observed density field. Finding
this displacement field corresponds to finding the solution of 
a Monge-Amp\`ere equation, or to solve a Monge-Kantorovitch
problem \citep{Brenier03}. This problem, once discretized,
is equivalent to searching for the minimum of
the discretized action
\begin{equation}
S=\sum_{i=1}^N\left({\bf q}_{j(i)}-{\bf x}_i\right)^2 .
\label{eq:mak-action}
\end{equation}
which assigns initial comoving Lagrangian positions ${\bf q}_j$ to final 
comoving Eulerian positions ${\bf x}_i$. 
We followed a different technique than in \cite{lavaux08} to account  for redshift distortions. Modifying the action using Zel'dovich approximation may lead to quite large errors for the reconstruction of the velocity of the observer. In a volume of radius $\sim$10\Mpch{}, the distortions are dominant compared to the Hubble flow, and introduce extra shell crossings as compared to real space coordinates. We use a technique inspired from earlier attempt using linear theory \citep{Pike05}. It consists in running an iterative algorithm on the smoothed  reconstructed peculiar velocity field. The steps are as follows:
\begin{itemize}
\item[(1)] We let $i$ be the iteration step number.
\item[(2)] We run a reconstruction of the peculiar velocities on a catalog \verb,Catalog(i),  of objects. The coordinates of these objects are assumed to be a good approximation of their real space coordinates. In that case we minimize the equation \eqref{eq:mak-action}.
\item[(3)] We smooth the reconstructed peculiar velocity in redshift space with a Gaussian kernel of radius 2.5\Mpch. This allows us to keep the large scale flow intact for the correction while alleviating the small scale effects for the correction of the distances of the objects of the catalog.
\item[(4)] We move the objects back from their redshift coordinates from catalog \verb,Catalog(0), to their presumed new real space coordinates using the velocity field obtained at the step (3). This new set of position allows us to create the catalog \verb,Catalog(i+1), of objects.
\item[(5)] We loop to point (1) until convergence.
\end{itemize}
We set \verb,Catalog(0), to be the 2MRS catalog in redshift coordinates. This algorithm in practice converges very quickly to the the true peculiar velocity field without redshift distortions, typically in 2 or 3 iterations. This effectiveness counterbalances the huge cost of running several MAK reconstruction. 

\section{Results: I. 2MRS velocity field and comparison with measured distances: estimation of $\Omega_\text{m}$}
\label{sec:result1}

In reconstructing the velocity field using MAK based on priors, we
first set $\Omega_m=0.258$ and bias \citep{Kaiser89} to 1 as indicated
by WMAP5 results \citep{WMAP5_LCDM} and using 2dF and SDSS/WMAP
results \citep{Tegmark2004,2df_cosmo}.  We then do a self-consistency
check on the presumed cosmological parameters by confronted
reconstructed and observed Local Group motion. This is the approach
taken in the next Section (Section~\ref{sec:result2}).  The velocity
of each galaxy in 2MRS is reconstructed using these parameters and
checked against observed distances within 3,000~\kms{}. This test is
detailed in Appendix~\ref{app:loc_bulk}.

In this Section, we take a different approach and leave $\Omega_m$
free and then constrain its value by maximizing the correlation
between the reconstructed and observed peculiar velocities. This
approach allows us to constrain the value of the bias parameter. The
2MRS velocities are reconstructed using a uniform grid of size $130^3$
sampling a cubic volume of $260^3 h^{-3}\text{Mpc}^3$ as shown in
Fig.~\ref{fig:2mrs_velfield}. The motion of the Local Group is
obtained using an interpolation based on the adaptive weighting of the
peculiar velocities of the objects that lie within 4-5\Mpch{} from us
(method detailed in Appendix~\ref{app:weigh}).  We have checked that
increasing the reconstruction resolution does not significantly change
the reconstructed velocities.

\begin{figure}
 \begin{center}
   \includegraphics[width=\hsize]{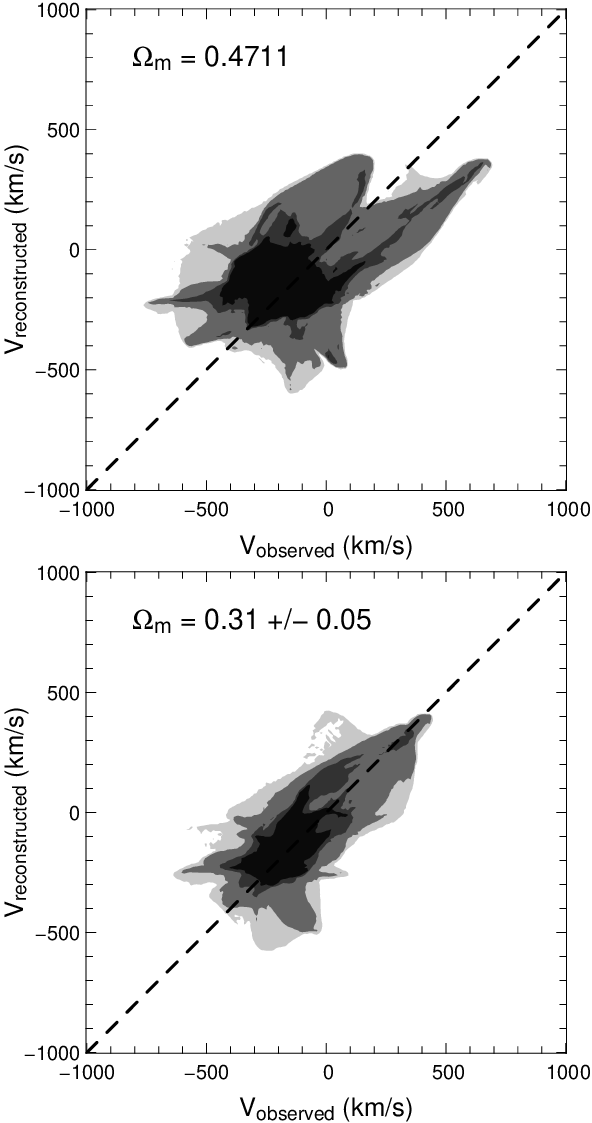}
 \end{center}
 \caption{\label{fig:velocity_comparison} {\it Observed
     vs. Reconstructed velocity field} -- These two panels give the
   result of the comparison between the reconstructed and the
   observed line-of-sight component of the velocity field. The top
   (respectively bottom) panel indicates the result of a comparison
   before (respectively after) having corrected for bulk
   flows. We have used the method of moments using the part of the
   scatter within the 95\% probability contour in both cases. For the
   top panel we obtain $\Omega_\text{m} \sim 0.47$. For the bottom
   panel, we estimate $\Omega_\text{m} \sim 0.31$. The correlation
   coefficient $r$ in that case is equal to 0.73, with a typical a posteriori
   velocity error of 65~\kms. To plot these two distributions, we
   assumed $h = 0.80$. Though the centroid of the scatter has been
   repositioned for the estimation of the slope, this correction has
   not been applied for the above distributions. Decreasing $h$
   would move the scatter to the right of the diagonal. The
   filled contours limit the regions that corresponds to 50\%
   (black), 68\% (dark gray), 95\% (gray) and 99\% (light gray) of
   joint probability to have a reconstructed and an observed velocity
   at some position in the Local Universe.}
\end{figure}

We present, in Fig.~\ref{fig:velocity_comparison}, the result of the
comparison of observed peculiar velocity field vs. reconstructed peculiar
velocity field in the volume of radius 3,000~\kms{}. Both fields have been
obtained using adaptive smoothed interpolation on the the
 line-of-sight component of the velocities (observed or reconstructed) of the
 objects put at their redshift position
 (Appendix~\ref{app:weigh}).  As we are using the redshift
 coordinates, and not the distance-induced coordinates, we should be free of
 the so-called volume Malmquist bias. Moreover, the two fields are
 enforced to be smoothed in exactly the same way at each spatial location,
  the one needing a larger smoothing scale prevailing on the
 other. Finally, we opted for smoothed interpolation considering that
 galaxies are fair tracers of the underlying continuous velocity field. 
As it is adaptive, the smoothing scale is left free, while in practice
it remains in the range [3.4,8.7]\Mpch{}.\footnote{This range corresponds to the selected smoothing radius for 95\% of the volume.}
The top panel gives the result of the comparison  
of the raw velocities. However, it may happen
that the bulk flow of the 3,000~\kms{} volume is badly reconstructed.
To avoid this problem, we subtract the reconstructed
(observed respectively) bulk velocity of the 3,000~\kms{} region from
reconstructed (observed respectively) peculiar velocities. The
comparison of the resulting fields is given in the bottom panel of
Fig.~\ref{fig:velocity_comparison}. The observed and reconstructed
velocities are now visually well-correlated. To test the correlation,
we use two quantities already introduced in \cite{lavaux08}. First,
we define the correlation coefficient
\begin{equation}
  r = \frac{\langle \tilde{v}_\text{rec,r} \tilde{v}_\text{obs,r}\rangle}{\sqrt{\langle \tilde{v}^2_\text{rec,r}\rangle \langle \tilde{v}^2_\text{obs,r} \rangle}}
\end{equation}
with $\tilde{v}_\text{rec,r} = v_\text{rec,r} - \langle v_\text{rec,r}
\rangle$ and $\tilde{v}_\text{obs,r} = v_\text{obs,r} - \langle
v_\text{obs,r} \rangle$, $v_\text{rec,r}$ the line-of-sight component
of the reconstructed velocity field, $v_\text{obs,r}$ the
line-of-sight component of the observed velocity field.  Second, we
define the typical reconstruction error
\begin{equation}
 \sigma^2 = \frac{1}{2} \langle\left(\tilde{v}_\text{rec,r} - \tilde{v}_\text{obs,r}\right)^2 \rangle
\end{equation}
We obtain $r = 0.73$ and $\sigma = 65$~\kms. 
These values shows that our reconstructed
velocity field is of good quality. Indeed, the standard deviation $\sigma$ is rather small compared to the 
typical extent of the PDF ($\sim$400\kms). The correlation $r$ is better than what has been
obtained for mock catalogs for which observational effects has been included ($r_\text{mock}\simeq 0.53$), in spite that the comparison is done on a smaller number of tracer of the peculiar velocity field: 576 tracers within 3,000~\kms{} in the case of this paper against $\sim$1600 in mock catalogs).
This seems to indicate that our results should be as good as the one obtained on simulations \citep{lavaux08}.

From the comparison between reconstructed to observed peculiar
velocities, it is possible to measure $\Omega_\text{m}$. Though a
pure likelihood approach would seem the natural way, \cite{lavaux08}
showed the high sensitivity to the prior of a likelihood which does
not take into account correlations of the velocity field. For this
paper, we will thus use the apparently less sensitive method of
moments \citep{VelGrav07,lavaux08} to determine $\Omega_\text{m}$ on
the adaptively smoothed velocity field. Smoothing has the advantage
of using the correlation of the velocity field to increase the
signal-to-noise. The noise coming from both errors on observed
distances and from MAK modeling will be in particular diminished.  As
it has been shown in this paper that the slope between reconstructed
displacements and observed peculiar velocities
\begin{equation}
  s = \sqrt{\frac{\langle \tilde{v}^2_\text{rec,r}\rangle}{\langle
      \tilde{v}^2_\text{obs,v} \rangle}}
\end{equation}
seems not to be statistically biased, we will use it here as our
estimator of $\Omega_\text{m}$.  We improve the technique by computing
a set of different slopes from data points that lies farther and
farther from the perfect correlation (i.e. the diagonal in
Fig.~\ref{fig:velocity_comparison}). This allows us to make an
estimate of the error bar on our result on $\Omega_\text{m}$ through
the use of the two other estimator of the slope $r \times s$ and $s /
r$ (Appendix B of \citeauthor{lavaux08} \citeyear{lavaux08}).

We assumed $h = 0.80$, a value
compatible with the catalog of distances with the current calibration
of distance indicators.  The value assumed for $h$ only
marginally affects our estimation of $\Omega_\text{m}$ as it was shown
in \cite{lavaux08}. From the 576 velocities that were reconstructed,
we estimate $\Omega_\text{m}= 0.31 \pm 0.05$. This result is in agreement with
\cite{Pike05}. This is only consistent at the lower end of $\Omega_\text{m}$ with previous results based on orbit reconstruction methods \citep[e.g][]{MohTu2005}. An extensive
likelihood analysis including correlations, which would yield a better
estimation, is postponed to another paper. 

We estimate a systematic error of $\sim 9\%$ on reconstructed peculiar
velocities due to the assumed values of cosmological parameters. To
this error, we add in quadrature a random reconstruction error of
70~\kms{} according to the mean (both on amplitude and by component), 
as estimated using reconstruction on simulations in redshift space (Fig.~\ref{fig:mockobservers}, Appendix~\ref{app:lin_vs_mak}).

One must be aware of
the danger of cosmic variance, as it has been highlighted in
\cite{lavaux08}. A way of checking it is to use the density field of
the whole 2MRS as an indicator of the local density fluctuation. We
computed a density contrast $\rho_{30}/\rho_{100} - 1 = -0.046$ for the
Universe within 30\Mpch{}, compared to the mean density within
100\Mpch{} using the mass density field obtained from 2MRS data. It
means the volume on which we make the comparison is slightly
underdense and thus may bias our value to lower
$\Omega_\text{m}$. However, the amount of the systematic bias due to
cosmic variance seems to be here negligible compared to errors due
solely to the noise.  Nevertheless, this value is in good agreement
with other measurements such as the one given by WMAP
\citep{WMAP5_LCDM} for the \LCDM{} model.  Having obtained and tested
our reconstructed velocities within 3,000~\kms{}, we go to larger
scales and study the origin of the motion of the Local Group with respect to CMB,
$V_{\rm LG/CMB}$, in the next section.


\section{Results: II. 2MRS velocity field and the origin of the CMB dipole}
\label{sec:result2}

\begin{figure}[t]
 \includegraphics*[width=\hsize]{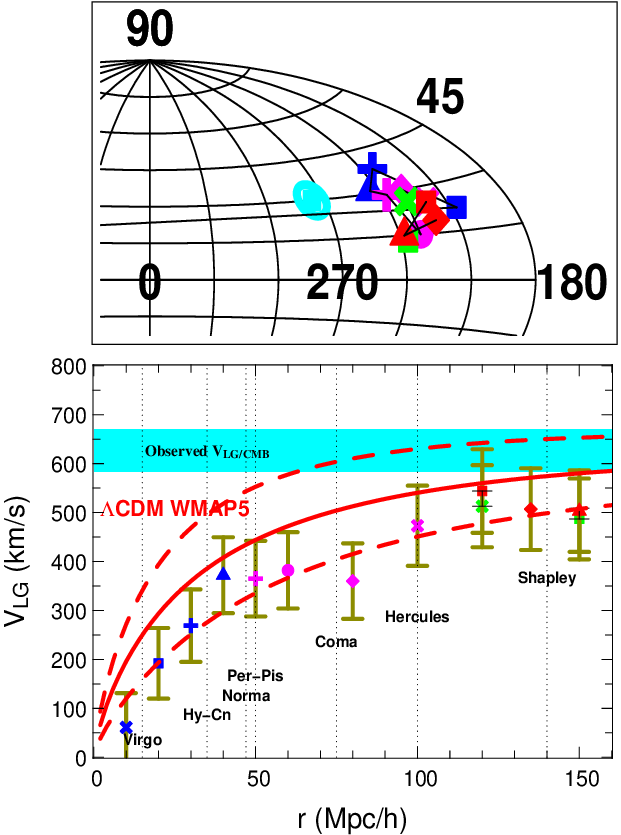}
 \caption{
{\it Velocity of the Local Group in progressively larger rest frames,}$V_{\rm LG}$:
The observed amplitude and direction of the CMB dipole motion are
shown by the horizontal cyan band in the lower panel and the solid
cyan large square in the top panel, respectively.  The lower panel
shows the amplitude of the velocity of the Local Group in successively larger rest
frames as the velocity field of 2MRS is reconstructed at increasing
radii.  Incompleteness is illustrated by the black error bar on data
points beyond 120\Mpch{} in the bottom panel.  The top panel shows how
the direction changes as the radius increases. The red curves in the
bottom panel indicate the prediction of growth of the velocity of the Local Group
for a WMAP5 cosmology.
The solid curve gives the
expectation of the reconstructed velocity for a survey whose radius is
indicated by the X axis. The two dashed curve indicates the $1\sigma$
fluctuation relative to the expectation given by the model. To compute these
curves, we used the WMAP5 parameters: the density of cold dark matter 
$\Omega_\text{c} = 0.212$, the density of baryons 
$\Omega_\text{b} = 0.044$, $h = 0.719$, $\sigma_8 = 0.77$ and a
\cite{EisensteinHu98} power spectrum (without baryonic wiggles).
\label{fig:dipole}}
\end{figure}

The CMB dipole motion, is obtained by using the reconstructed 3d
velocity field{ } generated using the 2MASS Redshift Survey.
The velocities are reconstructed within increasing
radii centered on the Local Group. A table of the resulting dipole as a
function of the reconstruction radius is given in
Table~\ref{table:dipole}.  The entries are represented in
Fig.~\ref{fig:dipole}.  No convergence is achieved below 120\Mpch. The
2MRS sample becomes highly incomplete beyond this scale and further
conclusions cannot be made before a more complete sample becomes
available. Whether the Shapley supercluster yields the convergence to
the CMB dipole also remains questionable, as farther away in the
southern hemisphere, dominant structures such as the
Horologium-Reticulum supercluster could change the direction of the
dipole. Thus, it is possible that the depth of the convergence towards
CMB dipole lies well-beyond the Shapley concentration itself.

\begin{figure}
 \includegraphics*[width=\hsize]{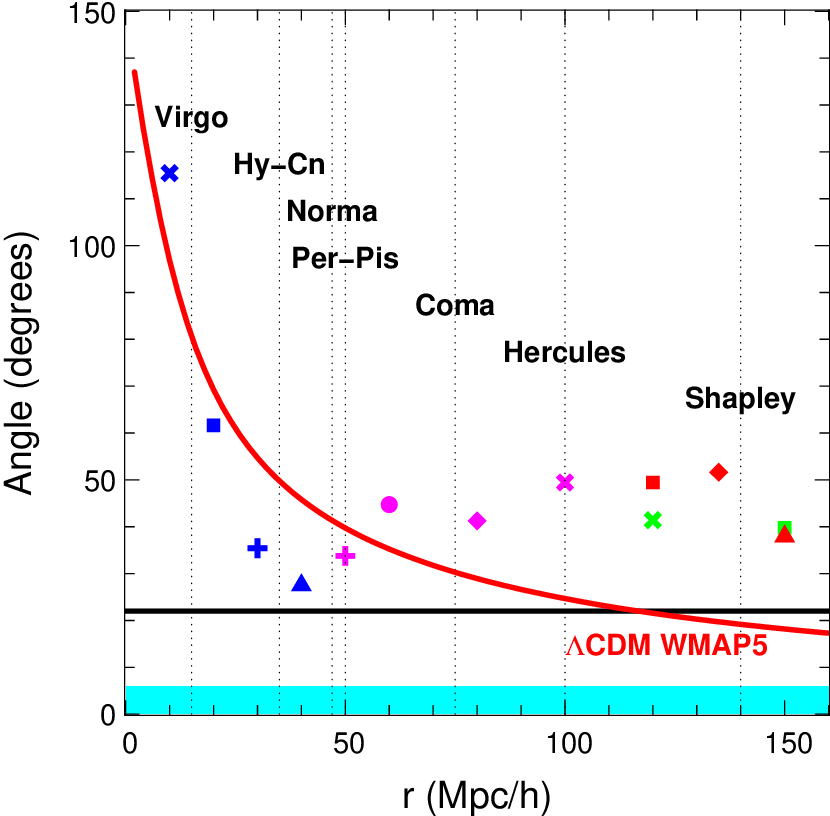}
 \caption{
   {\it Misalignment angle of the reconstructed Local Group velocity}
   -- We have plotted here the misalignment angle between the
   reconstructed Local Group velocity and the direction indicated by
   the CMB dipole. The symbols and colors that are used here are the
   same as in Fig.~\ref{fig:dipole}. The red curve represents the
   95\% probability limit of the misalignment for a \LCDM{} universe
   whose parameters have been chosen as estimated by WMAP5. 
   The
   horizontal black thick line gives the expected misalignment, at
   95\% of probability, between the reconstructed velocity and the
   observed motion of the Local Group. It has been estimated by
   applying the reconstruction to one \LCDM{} simulation.
   \label{fig:dipoleAngle}
 }
\end{figure}

We also plot the increase in the bulk flow relative to the CMB dipole over the reconstructed peculiar velocities within 120\Mpch{}
in the bottom panel of Fig.~\ref{fig:2mrs_velfield}.  The figure
demonstrates that less than half of the amplitude of the dipole is
generated within the volume enclosing Hydra-Centaurus-Norma. To reach convergence a
significant contribution to the dipole has to be made from the
Shapley supercluster and beyond. This agrees with an analysis of the
X-ray data \citep{Kocevski2006}.  We also fail to observe convergence
by the time we reach 120\Mpch.  Indeed, although the amplitude of the
reconstructed dipole seems to approach that of CMB, its direction
remains well away from it.  Using observational error bars, we
estimate that the observational error on the direction of the Local Group
velocity should not exceed $\sim 6^\circ$ at 95\% confidence. In addition,
we estimated from an $N$-body \LCDM{} simulation that there is an intrinsic
$22^\circ$ error (95\% confidence) in the direction of the
reconstructed velocity because of the modeling errors of the MAK
reconstruction. In sum, we expect the observed and reconstructed vectors to 
agree to within $\sim 23^\circ$ whereas they lie at $\sim 40-50^\circ$ from each
other. This problem is highlighted by Fig.~\ref{fig:dipoleAngle},
where we have represented the angular separation between the
reconstructed Local Group velocity and the CMB dipole. We note that
beyond $\sim 60$\Mpch{} the separation remains at roughly $40-50^\circ$.  We
now proceed to compare our results to predictions given by the
\LCDM{} model.

\section{Compatibility with the \LCDM{} model}
\label{sec:compatibility}

The measurements of the convergence of the velocity of the Local Group
to the CMB dipole can be used to check the \LCDM{} model. Indeed, the
statistic of peaks and voids in the density field imprints the
fluctuation and the rate at which the velocity of the Local Group
converges to its expected final value, which should be the one given by
the CMB according to the current cosmological paradigm. We will use
here the cosmological parameters estimated by WMAP5 \citep{WMAP5_LCDM}
and weak lensing measurements \citep{Fu08,Benjamin07,Massey07}. We use
a statistical modeling of the Local Group velocity growth developed by
\cite{Juszkiewicz89} (equation 12) and \cite{Lahav90}. The prediction is based on linear
theory and is able to give the probability of occurrence of our
measurements given the cosmological model. As proposed by
\cite{WMAP5_LCDM} and \cite{Fu08}, we use $\Omega_\text{m} = 0.256$
(the density of baryon is $\Omega_\text{b}=0.044$ and the density of
cold dark matter is $\Omega_\text{c}=0.212$) and $\sigma_8 = 0.77$. We
also set the Hubble constant to $H = 71.9$~\kmsMpc{} and the spectral
index $n=1$. In addition to the cosmology specified above, we enforce
that the amplitude of the velocity of the Local Group is 627~\kms. 

This conditioning
allows us to avoid the problem of the probability of our own velocity
in the different universes we consider in this work. Thus, though
we may be at a special point in the Universe, our results should not
be affected by the fact that the velocity of Local Group may be unusual. 
In the following, we thus focus on the conditional
probability of obtaining a set of reconstructed velocities of the local group $\{{\bf v}_i\}$ 
given a specification of the velocity of the Local Group ${\bf V}_\text{LG}$. 
Nevertheless, we have also checked in 
Section~\ref{sec:param_estimate} that
using the joint probability of having both $\{ {\bf v}_i \}$ and ${\bf V}_\text{LG}$ 
only slightly changes the probability contours in the specific case of our velocity
as given in Table~\ref{table:dipole}. This indicates that the velocity of
the Local Group is not too uncommon in universes whose cosmological parameters are
selected by the likelihood analysis.

We use a \LCDM power spectrum as given by \cite{EisensteinHu98} but
without incorporating baryonic wiggles. We have checked that
introducing wiggles does not change the prediction much though the
introduction of baryons does decrease the expectation of the
reconstructed Local Group velocity for distances $\la$60\Mpch.  This
behavior is expected as baryons tend to suppress density fluctuations
below the sound horizon while they are linked to photons by the mean
of the Compton effect.  The size of the horizon at the moment when
baryons separate from photons is typically $\sim 45 h^{-1}$~Mpc
\citep{EisensteinHu98}, which is the same scale at which we observe
a change due to the introduction of baryons. In the presence of
baryons the density field has less power on smaller scales, so it is more
difficult for the Local Group to acquire its velocity using only small
scale fluctuations. Fully computing the expected value of the Local
Group velocity for a given survey depth, we indeed observe that it
decreases by 5-15\% when we take into account baryons in the power
spectrum. 

Now, we may also consider the effect of changing $\sigma_8$. Its
principal effect is to change the amount of fluctuation of the
velocity field around its expected value. A growth of convergence that
is slow and regular corresponds most likely to a low local $\sigma_8$,
whereas a growth with a lot of independent fluctuations favors a high
$\sigma_8$. Its impact on the expectation of the amplitude of the
velocity field is more complicated. Indeed, cosmologies with high
$\sigma_8$ tends to have stronger fluctuations relative to the
expected velocity, which yields an higher expected amplitude. Thus
higher $\sigma_8$ should increase slightly the expectation of the
amplitude. This means that even if we use only the evolution of the
amplitude of the Local Group we should be sensitive to $\sigma_8$,
though more marginally than on the shape of the power spectrum
$\Omega_\text{m} h$. The impact of $\sigma_8$ is however dominant
concerning the fluctuations of the direction of the velocity of the
Local Group. Only universes with a high $\sigma_8$ allow this
direction to depart significantly from the one given by the CMB dipole
at scales larger than 60\Mpch{} as we will see in the section~\ref{sec:compat_wmap5}.

$\Omega_\text{m}$ represents the true dynamical content of the
universe. For a given realization of density fluctuations, the
dynamics is faster for a high $\Omega_\text{m}$ than for a low
$\Omega_\text{m}$. Thus the convergence of the velocity of the Local
Group should be quicker in a universe with a high $\Omega_\text{m}$
than in another one with a low $\Omega_\text{m}$. But the expectation
of this convergence does not tell anything about the probability on
the velocity of the Local Group itself.  We have decided in this work
not to presume anything on the probability of occurrence of the
velocity of the Local Group but only to check the consistency of the
dynamics of the Local Universe.

We are now going to use the prediction of the growth of the
Local Group velocity as both a checking of the \LCDM{} model and a way
to estimate cosmological parameters. We will explore a limited
number of parameters: the total mean density of matter
$\Omega_\text{m}$, the mean baryon density $\Omega_\text{b}$, the Hubble constant $H$ and the
variance $\sigma_8$ of density fluctuations within a sphere of radius
8\Mpch. 

In Section~\ref{sec:compat_wmap5}, we compare our results to predictions using
WMAP5 cosmological parameters. In Section~\ref{sec:param_estimate}, we estimate
cosmological parameters using both our measurements and the constraint coming from
nucleosynthesis.

\subsection{Comparison to WMAP5 results}
\label{sec:compat_wmap5}

First, we assume the \LCDM{} model described above and we obtain the
red curves represented in the bottom panel of
Fig.~\ref{fig:dipole}. The solid curve gives the expectation of the
convergence of the dipole given our motion. The dashed curves give the
estimated 1$\sigma$ fluctuation for the expectation of the velocity of
the Local Group for a given survey depth. This expectation is computed
{\it assuming} that the Local Group is moving at 627~\kms{} relative to
the CMB, as determined using CMB experiments.  Though it seems
astonishing to be systematically below the solid curve such a
behavior may be explained by the fact that we are dealing with
correlated measurements (see Section~\ref{sec:param_estimate}). For
each point, the \LCDM{} model seems to be in agreement with the Local
Group velocity reconstructed by MAK at the $1\sigma$ level. A similar
prediction can be obtained for the maximum angle separation between
the CMB dipole and the reconstructed velocity (equation 13).  This
prediction is represented by the solid red and thick curve in
Fig.~\ref{fig:dipoleAngle}. For a given survey depth, the misalignment
has a 95\% probability to be located below the curve. This is in
perfect agreement with the results given by the MAK reconstruction
until a depth of 60\Mpch{}. There, the misalignment goes up to $\sim
50^\circ$ and stays there. At 100\Mpch, we have a probability less
than $3\times 10^{-5}$ that the misalignment is greater than
50$^\circ$.

Given the importance of the disagreement between the predicted maximum
misalignment and the observed one, we see three
options. First, the reconstructed velocity of the
  Local Group may be exceptionally plagued by redshift distortions,
  even though on the rest of the volume the reconstructed velocity
  field is very well reconstructed. In this case, we are in the same
  position as number-weighed methods, which like in \cite{Erdogdu2005}
  yield a large misalignment. Doing a complete Monte-Carlo analysis on
  mock catalogs of these problems are beyond the scope of this paper,
  and we postpone this to future work.  Second, we may miss some
relatively important structures in the masked region. We remind that
due to severe incompleteness we are missing information on structures
below 5$^\circ$ for $|l| > 30^\circ$
and 10$^\circ$ for $|l| < 30^\circ$.  One of the brightest X-ray
clusters of the Local Universe is the Ophiuchus Cluster at 8400~\kms{}
and it may well reside in a hidden supercluster of significance
\citep{Wakamatsu2005}. Second, more dramatically, it would mean that
large bulk flows exist at a scale of 100\Mpch, typically $\sim
500$~\kms. This is extremely unlikely in a \LCDM{} cosmology as it is
already highlighted by Fig.~\ref{fig:dipoleAngle}. A similar result
has been obtained recently by \cite{Feldman08} and \cite{Watkins08}
but using observed velocity fields. On the contrast, using linear theory
on 2MRS \citep{Erdogdu2005,Erdogdu2006} gives a low value of bulk flow
and disagrees with \cite{Feldman08} and \cite{Watkins08}.

\subsection{Parameter estimation}
\label{sec:param_estimate}

From the measurements of the growth of the Local Group velocity, it is
possible to estimate the joint probability of $\Omega_\text{m}$ and
$\sigma_8$ by a maximum likelihood approach. To do that, we use a
likelihood analysis on the velocity of the Local Group in the
different rest frames that we obtained using 2MRS in
Fig.~\ref{fig:dipole}. A similar analysis was
first introduced by \cite{Kaiser89} to compare predicted dipoles to
observed bulk flows. Here, we focus on the expected evolution of the
bulk flow on different scales and how it should compare to our own
reconstructed bulk flow in a given cosmology.  We detail, in
Section~\ref{sec:stat_growth}, the statistics of the growth of the
Local Group velocity using linear theory, an idea  
originally introduced by \cite{Juszkiewicz89}, \cite{Lahav90} and \cite{Strauss92}. In
Section~\ref{sec:stat_with_err}, we see how to mix observational
errors into this statistical modeling. Finally, in
Section~\ref{sec:stat_results}, we see the results of this
analysis on the whole set of data points that we obtained on the value
of the cosmological parameters.

\subsubsection{Statistics of the growth of the Local Group velocity}
\label{sec:stat_growth}

To estimate the likelihood of the cosmological parameter vector $p$,
we need the conditional probability $P({\bf v}_1,{\bf v}_2,\ldots,{\bf
 v}_N|{\bf V}_\text{LG},p)$ of reconstructing the velocities of the
Local Group $\{{\bf v}_i\}$ in the set of rest frames $\{\mathcal{R}_i\}$
given that the true velocity of the Local Group is ${\bf V}_\text{LG}$
and the cosmological parameters are described by $p$. To estimate this
probability, we will follow \cite{Juszkiewicz89} and use linear theory
and assume the components of the velocity field are independent
Gaussian random variables. To ease the notation, we write
$\mathfrak{U}_k$ to designate the $N$ dimensional vector whose $i$-th
component is the $k$-th component of the $i$-th vector:
$\mathfrak{U}_{k,i} = ({\bf v}_i)_k$. Using linear theory, we can
compute the variance $\sigma_i$ of the velocity of the Local Group
computed using a homogeneously sampled spherical survey of depth $R_i$
\begin{equation}
 \sigma^2_i = \frac{(\beta H)^2}{6\pi^2} \int_{k=0}^{+\infty} P(k) \widetilde{W}^2(k R_i)\text{d} k
\end{equation}
with $\widetilde{W}(x) = 1 - \sin(x)/x$. We also define
\begin{equation}
  \sigma^2_V = \frac{(\beta H)^2}{6\pi^2} \int_{k=0}^{+\infty} P(k) \text{d}k
\end{equation}
the variance of the velocity field.
Similarly, one may obtain the correlation coefficient $\gamma_{i,j}$
between the component $k$ of ${\bf v}_{i}$ and ${\bf v}_{j}$
\begin{multline}
 \gamma_{i,j} = \frac{1}{\sigma_i\sigma_j} \langle v_{k,i} v_{k,j} \rangle = \\
 \frac{(\beta H)^2}{6 \pi^2 \sigma_i\sigma_j} \int_{k=0}^{+\infty} P(k) \widetilde{W}(k R_i) \widetilde{W}(k R_j)\,\text{d} k
\end{multline}
We also need the correlation $\Gamma_i$ between $v_{k,i}$ and $V_{k,\text{LG}}$
\begin{multline}
 \Gamma_{i} = \frac{1}{\sigma_i \sigma_V} \langle v_{k,i} V_{k,\text{LG}} \rangle =\\
\frac{(\beta H)^2}{6 \pi^2 \sigma_i \sigma_V} \int_{k=0}^{+\infty} P(k) \widetilde{W}(k R_i)\,\text{d} k\,.
\end{multline}
We note that, even in the non-linear regime, the velocity field
remains highly Gaussian. However, these coefficients
do not take into account potential distortions to its correlations due
to the MAK reconstruction and non-linear effects. Though, we expect
the computation of these coefficients to be correct on large scales,
precise measurements using $N$-body simulations would be required here
to improve the precision on the final estimation of cosmological
parameters when smaller scales are used. This measurement could be interesting as small scales may
contain more information on $\Omega_b$ and $\sigma_8$.
Using these quantities it is now possible to compute the
probability $P({\bf v}_1,{\bf v}_2,\ldots,{\bf v}_N|{\bf V}_\text{LG})
=
P(\overrightarrow{\mathfrak{U}}|{\bf V}_\text{LG})$, with $\overrightarrow{\mathfrak{U}} = (\mathfrak{U}_1,\mathfrak{U}_2,\mathfrak{U}_3)$. 
We refer the reader to the appendix for the details of the computation. 
We carry here the result given in
equation~\eqref{eq:conditional_gaussian}:
\begin{multline}
 P(\mathfrak{U}_k|V_{k,\text{LG}}) = \frac{\sqrt{\det M_s}}{(2\pi)^{N/2}} \prod_i \sigma_i^{-1}  \\
 \times \exp\left(-\frac{1}{2} \sum_{i,j=1}^N \mathfrak{W}_{k,i} \mathfrak{W}_{k,j} M_{s,i,j}\right) \label{eq:likelihood_pure_one_component}
\end{multline}
with
\begin{equation}
 \mathfrak{W}_{k,i} = \frac{\mathfrak{U}_{k,i}}{\sigma_i}-\frac{\Gamma_i V_{k,\text{LG}}}{\sigma_{V}}\,.
\end{equation}
and $M_s$ the top-left part of the invert of the covariance
  matrix as defined in appendix by Eq.~\eqref{eq:covariance_matrix}.
As the three components of the Local Group velocity are independent,
the total probability of measuring the tensor $\overrightarrow{\mathfrak{U}}$
given that the velocity of the Local Group is ${\bf
 V}_\text{LG}$ is thus:
\begin{equation}
 P\left(\overrightarrow{\mathfrak{U}}|{\bf V}_\text{LG}\right) = 
 \prod_{k=1}^3 P\left(\mathfrak{U}_k|V_{k,\text{LG}}\right)\,. \label{eq:likelihood_pure_three_components}
\end{equation}
However, in Fig.~\ref{fig:dipole}, we see that errors coming from
reconstructions are quite significant. We thus have to take them
properly into account into the likelihood analysis.

\begin{widetext}

\subsubsection{Measurement errors}
\label{sec:stat_with_err}

To account for reconstruction errors, we assume that they are
independent from one survey depth to another and they are properly
modeled by a Gaussian distribution. We will thus write the probability
that the component $k$ of the reconstructed velocity ${\bf
 v}_\text{\text{rec},i}$ of the Local Group for a survey cut $i$, given
that the true velocity should be ${\bf v}_{i}$ for some cosmological
model and the expected error dispersion for the component $k$ is
${\sigma}_{e,k,i}$,
\begin{equation}
 P(v_{\text{rec},k,i}|v_{k,i},\sigma_{e,k,i}) = \frac{1}{(2\pi)^{1/2} \sigma_{e,k,i}} \exp\left(\frac{1}{2} \left(\frac{v_{\text{rec},k,i} - v_{i}}{\sigma_{e,k,i}}\right)^2\right) \label{eq:noise1d}
\end{equation}
where $v_{\text{rec},k,i}=({\bf v}_{\text{rec},i})_k$ and
$v_{k,i}=({\bf v}_i)_k$.  To ease the notation, we use now the
$N$-dimensional vector $\mathfrak{V}_k$, whose components are
$\mathfrak{V}_{i,k}=({\bf v}_{\text{rec},i})_k$. So these vectors holds the
component $k$ of the reconstructed velocities for all survey depth. 

We may now merge the two probabilities
\eqref{eq:likelihood_pure_one_component} and \eqref{eq:noise1d} to
obtain the total probability to measure $\mathfrak{V}_k$ given a
cosmology determined by the vector of parameters $p$ and the
component $k$ of the velocity of the Local Group $V_{\text{LG},k}$:

\begin{equation}
 P(\mathfrak{V}_{k}|V_{\text{LG},k},p) = 
 \int_{\bf U} \text{d}{\bf U} \left(\prod_{i=1}^N P(v_{\text{LG},i,k}|U_i,\sigma_{e,i,k})\right)  
 P(U_1, U_2, \ldots, U_N|V_{\text{LG},k},p)\,,
\end{equation}
with $\mathbf{U}$ a vector with $N$ components.  One may use a similar
transformation as used in \eqref{eq:exp_argument} to compute the above
integral and obtain the sought above probability:
\begin{equation}
 P(\mathfrak{V}_{k}|V_{\text{LG},k},\mathfrak{S}_{e,k},p) = 
 \frac{1}{(2\pi)^{N/2}} \sqrt{\frac{\det M_s}{\det M_{\text{obs}}(\mathfrak{S}_{e,k})}} \exp\left(-\frac{1}{2} f(\mathfrak{V}_{k}, V_\text{LG,k}, \mathfrak{S}_{e,k})\right) \label{eq:proba_full}
\end{equation}
where we defined the following quantities
\begin{eqnarray}
 f(\mathfrak{V},V,\mathfrak{S}) & = & \sum_{i=1}^N \left(\frac{\mathfrak{V}_i - U_{\text{mean},i}}{\mathfrak{S}_i}\right)^2 
 + \sum_{i,j=1}^{N} \mathfrak{W}_{\text{mean,j}} \mathfrak{W}_{\text{mean},i}  M_{s,i,j}\,, \\
 \mathfrak{W}_{\text{mean},i}(\mathfrak{V},V,\mathfrak{S}) & = & U_{\text{mean},i}(\mathfrak{V},V,\mathfrak{S})/\sigma_i - \Gamma_i V / \sigma_V\,, \\
 U_{\text{mean},i}(\mathfrak{V},V,\mathfrak{S}) & = & \sum_{j=1}^N [M(\mathfrak{S})]_{\text{obs},i,j}^{-1} \left(\frac{\mathfrak{V}_j}{\mathfrak{S}^2_{j}} + \frac{V}{\sigma_V} \sum_{k=1}^N M_{s,j,k} \Gamma_k \right)\,, \\
 M_{\text{obs},i,j}(\mathfrak{S}) & = & M_{s,i,j} + \frac{\sigma_i^2 \delta_{i,k}}{\mathfrak{S}^2_i}\,.
\end{eqnarray}
It is interesting to note that $U_{\text{mean},i}$ is a vector with
$N$ components which corresponds to the best average velocity given
the cosmological model and observations. In some sense, it is a
velocity  whose observational noise has been subtracted thanks to the modeling.  Finally, we
may assemble the statistics of the three independent variable
$\mathfrak{V}_k$ with $k=1,2,3$. We may do so by multiplying their
respective probability distribution given by Eq.~\eqref{eq:proba_full}
to obtain
\begin{equation}
 P({\bf v}_1,\ldots,{\bf v}_N|{\bf V}_\text{LG},\sigma_{e,1},\ldots,\sigma_{e,N},p) = \prod_{k=1}^N P(\mathfrak{V}_k|V_{\text{LG},k},\mathfrak{S}_k,p)\,.
\end{equation}
We now use this probability and the Bayes theorem to compute the probability
for the cosmological parameters $p$
\begin{equation}
 P(p|{\bf v}_1,\ldots,{\bf v}_N,{\bf V}_\text{LG},\sigma_{e,1},\ldots,\sigma_{e,N}) = \frac{P({\bf v}_1,\ldots,{\bf v}_N|{\bf V}_\text{LG},\sigma_{e,1},\ldots,\sigma_{e,N},p) P(p)}{\int_p \text{d}p\; P({\bf v}_1,\ldots,{\bf v}_N|{\bf V}_\text{LG},\sigma_{e,1},\ldots,\sigma_{e,N},p) P(p)} \label{eq:total_model_probability}
\end{equation}
with $P(p)$ a prior on the probability of $p$. In practice, we use a flat prior
on $\Omega_\text{m} h^2$ and $\sigma_8$  parameters and impose a prior on $\Omega_\text{b} h^2$ and $H$ as we detail in the next section.
\end{widetext}
\null

\subsubsection{Results}
\label{sec:stat_results}

\begin{figure*}
 \begin{center}
   \includegraphics[width=\hsize]{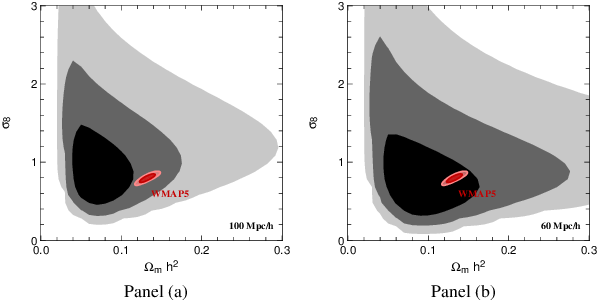}
 \end{center}
 \caption{\label{fig:sigma8_omegam} {\it Marginalized distribution of
     the joint probability of $\Omega_\text{m} h^2$/$\sigma_8$} -- In
   red, we represented the isocontour at 68\% (dark red) and 95\%
   (light red) of marginalized probability as given by WMAP5 (CMB
   data only). Left panel (a): We represented the isocontour at 68\% (dark
   gray) and 95\% of probability (light gray) as given by the
   likelihood analysis on the growth of the velocity of the Local
   Group up to 100\Mpch. For the gray contours, we introduced a prior on the
   mean density of baryons obtained by measuring the deuterium
   abundance \citep{Pettini08}, which gives $\Omega_\text{b} h^2 =
   0.0213 \pm 0.0010$. 
   Right panel (b): The dark (68\%) and light (95\%) gray contours} have been obtained in the same way
   as the gray contours of panel (a) except that we limited the depth of surveys to 60\Mpch{} inclusive.
\end{figure*}

We set here a Gaussian prior on the Hubble constant to $h=0.72\pm 0.08$ as specified by the HST Key project \citep{cepheid} and the spectral index to
$n_S=1$. Nonetheless, we checked that the result depends only weakly on this prior. 
We also fix the observed velocity of the Local Group to $V_\text{LG/CMB} =
627$~\kms. As we are using the power spectrum of \cite{EisensteinHu98}
without wiggles the cosmological parameter vector is thus
$p=(\Omega_\text{m} h^2,\Omega_\text{b} h^2,\sigma_8,
h)$. Note that the likelihood analysis, depending on the density power
  spectrum, should naturally depend on the shape $\Omega_\text{m} h$ and not on
  $\Omega_\text{m} h^2$. On the other hand, WMAP5 data yield $\Omega_\text{m}
  h^2$, so we arbitrarily chose WMAP5 parametrization over the natural one.
  The expected growth of the Local Group velocity
  presents a substantial degeneracy in the $(\Omega_m h^2,\Omega_b
  h^2)$ plane for scales smaller than the sound horizon ($\sim
  45$\Mpch). To get better constraints, we need to introduce some
  prior on one of these two variable. Following the results obtained
using the deuterium abundance \citep{Pettini08}, we set
a prior on $\Omega_\text{b} h^2$ to a Gaussian
distribution centered on $\omega_b = \Omega_\text{b} h^2 = 0.0213$ and
whose width is $0.0010$:
\begin{equation}
 P(p) \propto \exp\left(-\frac{1}{2}\left(\frac{\omega_b - 0.0213}{0.0010}\right)^2\right)\,,
\end{equation}
with a constant of proportionality independent of $p$. We are 
 not using the results from WMAP5 to make an independent comparison of their constraints and ours. We assume a
flat prior on $\omega_m = \Omega_\text{m} h^2$ and $\sigma_8$ though
we enforce them to be within the ranges $\omega_m \in [0,1.0]$ and
$\sigma_8 \in [0.05,7]$. 

The results given by Eq.~\eqref{eq:total_model_probability} are
illustrated in Fig.~\ref{fig:sigma8_omegam}. In the left panel, the 68\% (respectively
95\% and 99.7\%) isocontour of the marginalized joint probability of
$(\Omega_\text{m} h^2, \sigma_8)$ is represented in filled black
(respectively dark gray and light gray) contour when all velocities up
to 100\Mpch{} are taken into account. In the right panel, we represent
the isocontours of probability when one considers only velocities up to
60\Mpch. The 68\% (respectively 95\%) probability for cosmological parameters using
WMAP5 data alone are given by the hashed dark red (respectively hashed light red) filled contours in both panels.

It can be seen that our results in the right panel are compatible
within 1$\sigma$ with WMAP5, as far as $\omega_\text{m}$ and $\sigma_8$ are
concerned. However, in the left panel, the gray contours are significantly farther from WMAP5,
though still within a 2$\sigma$ limit. This behavior is expected as we
have already noticed that the misalignment angle does not seem to be
compatible with a standard WMAP5 cosmology, though 
the growth of the amplitude is compatible. The introduction of points beyond $\sim$60\Mpch{} shifts the
contours of probability to low $\omega_m$ and high $\sigma_8$. A high $\sigma_8$ allows
strong fluctuation in the direction, thus increasing the expected amplitude, whereas a low $\Omega_m$ limits the impact of these fluctuations on
the amplitude of the velocity by making the convergence slower. At the same time these contours become smaller as the constraints are enforced by a larger number of data points. The analysis of all data points up to 100\Mpch{} shows
that such a misalignment is less unlikely than what we would have
expected from the observation of the curve of
Fig.~\ref{fig:dipoleAngle}. This can be explained by the fact the
misalignment angles are not independent random variables. A more
independent set of random variables would be the difference between
the acceleration of the velocity the Local Group when one goes from one
  survey to a deeper survey.
Thus, the strong misalignments of $\sim 50^\circ$ for
all $R \geq 60$\Mpch, as we have on Fig.~\ref{fig:dipoleAngle}, might
be generated by a single rare event within $60-100$\Mpch{} from the observer 
which is not included in our present catalog, 
say a major supercluster in the Ophiuchus direction.
The relative absence of matter in our catalog compared to the reality would push away the reconstructed velocity from the CMB direction, which corresponds to what we observe. We need to
observe in greater detail the obscured part of the sky in these range
of distances to resolve this issue.
Another related reason for this misalignment might be
 an excessive mass in our model in the Perseus-Pisces
 supercluster. In the region delimited by a sphere of 
15\Mpch{} centered on z=5000~\kms, $l=140.2$, $b=-22$, we have
put $7.45\times 10^{15}\;\Msun$. According to \cite{Hanski01},
this is on the high end of the expected mass of the Perseus-Pisces
but not too excessive. A lower mass for this supercluster would
push the velocity of the Local Group in the direction of the CMB.

We may check the adequacy of the conditional probability $P(\overrightarrow{\mathfrak{V}}|{\bf V}_\text{LG},\overrightarrow{\mathfrak{S}},p)$
of reconstructing the set of Local Group velocities $\overrightarrow{\mathfrak{V}}$ given our observed motion ${\bf V}_\text{LG}$
with the motion of the Local Group itself.
To do that, we look at the total probability 
$P(\overrightarrow{\mathfrak{V}},{\bf V}_\text{LG}|\overrightarrow{\mathfrak{S}},p) = P(\overrightarrow{\mathfrak{V}}|{\bf V}_\text{LG},\overrightarrow{\mathfrak{S}},p) \times P({\bf V}_\text{LG}|p)$.
We checked that in this case our results are not significantly
affected, slightly moving the contours of both panels of Fig.~\ref{fig:sigma8_omegam}
towards higher $\omega_\text{m}$ and higher $\sigma_8$. This behavior
is expected. This probability takes into account the fact that it is
more likely to observe the velocity of the Local Group with respect to
CMB in a denser universe and with larger density
fluctuations. However, these considerations are insufficient to
significantly change the computed contours.

On the contrary, we may relax the condition on the
asymptotic reconstructed velocity, which is in principle given by
the observed CMB dipole. In that case, we rely only on the evolution
of the reconstructed velocities with the depth of the survey.  Such
an analysis shows the growth of the reconstructed velocity of the
Local Group is compatible within 1$\sigma$. However, this agreement
is mainly due to the extension of the contours of probabilities of Fig.~\ref{fig:sigma8_omegam} because
we do not enforce the observed motion of the Local Group. Doing so,
we becomes less sensitive to potential large scale bulk flows
and are unable to draw any meaningful conclusion about the cosmology.

To summarize, our results of this section indicate that, on one hand,
the total matter density might be lower in the local Universe than that
measured using WMAP5. On the other hand, the amount of fluctuation, 
$\sigma_8$, is more likely to be higher. Quantitatively, the maximum
 of the likelihood for the gray contours of the right panel is located
at $\omega_m \sim 0.08$ and $\sigma_8 \sim 0.57$.
The mean of the gray distribution of the right panel are given
by $\Omega_\text{m} h^2 = 0.11 \pm 0.06$ ($\Omega_\text{m} \sim 0.21 \pm 0.11$)
 and $\sigma_8 = 0.90 \pm 0.42$. 
For the gray contours of the left panel, this
position is shifted to $\Omega_\text{m} h^2 = 0.08 \pm 0.03$
($\Omega_\text{m} \sim 0.15 \pm 0.06$) and $\sigma_8 = 1.0 \pm 0.4$.
The most likely parameters are $\omega_\text{m} = 0.090$ and $\sigma_8 = 0.58$ for the right panel, and $\omega_\text{m} = 0.070$ and $\sigma_8 = 0.78$ for the left panel.
We note that putting our estimate of $\Omega_\text{m}$ of 
Section~\ref{sec:result1} and the above estimate of $\omega_\text{m}$, 
we may compute $h\simeq 0.51$, which is at odd with the estimate from the
HST Key project. We expect some overlap at the low $\omega_\text{m}$ and high $\sigma_8$ part of the contour with \cite{Watkins08}.
These discrepancies may be explained in three different
manners. First, some more work on simulations are required to understand the
potential systematics on these particular measures. Second, 
 we know that there exists a large concentration of galaxies in
the direction of the Ophiuchus at about 80\Mpch~\citep{Wakamatsu2005}.
This concentration is estimated to be as large as the Hercules 
supercluster and would need to be 
properly accounted in our statistical modeling as it is likely to 
systematically affect our reconstructed velocity. Third, we may live in a very
special part of the Universe and we would thus need to go to larger scales
using deeper catalogs to clarify this issue.

\section{Conclusion}
\label{sec:conclusion}

We have evaluated the 3-dimensional 2MRS peculiar velocity field using
a Lagrangian method (MAK) which works well into the nonlinear regime,
on scales above 4-5\Mpch{}. The method has been adapted to work
directly with redshifts. The reconstructed velocities are
well-correlated with the distances observed within a radius of 3000 km s$^{-1}$
(3k sample). The
reconstructed velocity of the Local Group in the rest frame of the 3k sample also
agrees well with the observed value.  In addition to this test, we
have successfully compared the observed velocity field to the
reconstructed ones within this same volume.
This comparison has lead to an estimate
of $\Omega_\text{m} = 0.31\pm 0.05$.

We have then studied the origin of the
Local Group motion in the CMB rest frame by using our 3d reconstructed
velocities. We have shown that less than half of the CMB dipole could
be generated within a radius enclosing the Hydra-Centaurus-Norma
supercluster. We have demonstrated how the trend of the convergence
varies up to 120\Mpch{} and have shown that convergence in position to
the CMB dipole is not reached even by this distance.

We checked that our measurements of the amplitude are qualitatively in
good agreement with the \LCDM{} model for cosmological parameters
given by WMAP5. We also note a weak dependence of the
  theoretical expectation of the amplitude on the presence of
baryons that could be of interest to get better constraints if future work
manages to reconstruct sufficiently well (at less than $\sim$5\% error)
the local dynamics. However, misalignment angles are significantly larger
than anticipated by the \LCDM{} model on scales larger than
50-60\Mpch.  To quantitatively check our measurements, we developed a
Bayesian analysis of the growth of the velocity of the Local Group in
the context of the linear theory but including correlations. We have
shown that the growth of the velocity of the Local Group may be a
powerful independent tool to explore cosmological parameters and of
the dynamics of the Local Universe.  This analysis yields an
independent measurement of $(\Omega_\text{m} h^2, \sigma_8)$ assuming
the quantity of baryons using the nucleosynthesis theory. For scales
up to 60\Mpch, our measurements agree with WMAP5 at the 1$\sigma$ level.
 For scales up to 100\Mpch, this agreement drops
to 1 to 2$\sigma$. This problem may have different origins:
the incompleteness of the 2MRS, bad correction of redshift distortions at
the introduction of the Perseus-Pisces or a real large bulk flow on the scale
of $\sim$100\Mpch, which can represent a challenge for \LCDM{}. 
More work on mock catalogs are needed to clarify this issue,
which is beyond the scope of this paper.

Forthcoming deeper and more complete redshift surveys especially those
in X-ray, could soon establish whether one has to go well beyond
120\Mpch{} to recover the CMB dipole or whether the Shapley
concentration at around 150\Mpch{} is sufficient to finally reach the
convergence. Such a larger survey would also allow a better
comparison of our analysis of observations with the ones made on
simulations on a similar scale to check \LCDM{} features in the large
scale velocity field, such as in \cite{StraussOstriker95}.

\section*{Acknowledgments}

We thank Joseph Silk for many enlightening suggestions. We thank the referee,
Michael Hudson, for important contributions.
We thank the 2MASS Redshift Survey
collaboration for having kindly provided the observational data without
which this study would not have been possible.
G.L. and R.M. thank Michael Hudson and Jeremiah Ostriker for very
helpful suggestions in particular on test of \LCDM{} model.
G.L., R.M. and B.T. acknowledge travel grants from French ANR
(OTARIE). We thank S. Prunet, M. Chodorowski, T. Sousbie, 
E. Komatsu and O. Lahav for useful discussions. 
This publication makes use of data products from the
Two Micron All Sky Survey, which is a joint project of the
University of Massachusetts and the Infrared Processing and Analysis
Center/California Institute of Technology, funded by the National
Aeronautics and Space Administration and the National Science
Foundation.
This research was supported in part by the National Science Foundation
through TeraGrid resources provided by the NCSA. TeraGrid systems are
hosted by Indiana University, LONI, NCAR, NCSA, NICS, ORNL, PSC,
Purdue University, SDSC, TACC and UC/ANL.

\newlength{\myextralen}
\setlength{\myextralen}{.9cm}

\begin{table*}
  \caption{\label{table:dipole} The reconstructed CMB dipole in progressively larger rest frames $R$}
  \begin{center}
    \begin{tabular*}{.85\hsize}{@{\extracolsep{\fill}}cccccccc}
      \multicolumn{8}{c}{
        \begin{tabular}{lp{.5\hsize}}
          Observed velocity: & $V_\text{LG/CMB}$ = 627$\pm$22~\kms, $l = 276 \pm 3$, $b = 30 \pm 3$
        \end{tabular}
      }\\
      \multicolumn{8}{c}{$v_{\text{LG/CMB},x} = 56 \pm 28$~\kms, $v_{\text{LG/CMB},y} = -540 \pm 25$~\kms, $v_{\text{LG/CMB},y} = 313 \pm 30$~\kms }
      \\
      \\
      \hline
      \hline
      \multirow{2}{1.5cm}{\hfill $R$ \hfill}  & \multicolumn{7}{c}{$V_\text{LG/R}$} \\
      \cline{2-8}
      & \hspace{\myextralen}$v_x$\hspace{\myextralen} & \hspace{\myextralen}$v_y$\hspace{\myextralen} & \hspace{\myextralen}$v_z$\hspace{\myextralen} & $|V|$ & $l$ & $b$ & angular \\
      (\Mpch) & (\kms) & (\kms) & (\kms) & (\kms) & (deg) & (deg) & separation \\
      \hline
      20   & $-157\pm 71$ & $-83 \pm 70$ & $73 \pm 70$   & 192 & 207 & 22 & 62\\
      30   & $-122\pm 71$ & $-173 \pm 72$ & $166 \pm 72$  & 269 & 234 & 38 & 35\\
      40   & $-137 \pm 71$ & $-285 \pm 74$ & $196 \pm 72$  & 372 & 244 & 31 & 27\\
      50   & $-172 \pm 72$ & $-270 \pm 74$ & $177 \pm 72$  & 365 & 237 & 29 & 34\\
      60   & $-236 \pm 73$ & $-282 \pm 74$ & $101 \pm 70$  & 382 & 230 & 15 & 45\\
      80   & $-208 \pm 72$ & $-230 \pm 73$ & $182 \pm 72$ & 360 & 228 & 30 & 41\\
      100  & $-328 \pm 76$ & $-277 \pm 74$ & $197 \pm 72$ & 473 & 220 & 25 & 49\\
      120  & $-378 \pm 78$ & $-319 \pm 76$ & $228 \pm 73$ & 544 & 220 & 25 & 49\\
      150  & $-257 \pm 74$ & $-413 \pm 79$ & $133 \pm 71$ & 504 & 238 & 15 & 38 \\
      \hline
    \end{tabular*}
  \end{center}
      {{\sc Note --} Glossary of the symbols used in the above tables. $l$
        and $b$ are the galactic longitude and latitude respectively. $R$
        is gives the radius of the sub-volume of the catalog on which the
        reconstruction is achieved. $v_i$ are the Cartesian galactic
        coordinates of the reconstructed velocity of the Local Group. The
        $x$ axis points towards $(l=0,b=0)$, the $y$ axis to
        $(l=90^\circ,b=0)$ and $z$ to $b=90^\circ$. 
      }
\end{table*}

\clearpage
\appendix
\section{Appendix A\\[.5cm] Conditional multivariate Gaussian random variables}

We consider $N+1$ Gaussian random variables described by the vector
$(\mathbf{U},V)$ with $\mathbf{U}$ a vector of dimension $N$ and $V$ a
scalar.  In this appendix, we compute the statistics of $\mathbf{U}$
given $V$.  We put the normalized variables $\widehat{U}_i = U_i /
\sigma_i$ and $\widehat{V} = V / \sigma_V$.  We assume that the
correlation coefficients $\gamma_{i,j} = \langle \widehat{U}_i
\widehat{U}_j \rangle$ between the components $U_i$ and $U_j$ of
$\mathbf{\widehat{U}}$ are given. We also assume that $\Gamma_i =
\langle \widehat{U}_i \widehat{V} \rangle$ is the correlation
coefficient between $\widehat{V}$ and $\widehat{U}_i$. The covariance
matrix of this $N+1$ variables may thus be written as
\begin{equation}
 C_{i,j} = 
 \begin{cases}
   1 & \text{if } i = j; \\
   \gamma_{i,j} & \text{if } i \leq N \text{ and } j \leq N\,;  \\
   \Gamma_{i} & \text{if } i \leq N \text{ and } j = N+1\,; \\
   \Gamma_j & \text{if } i = N+1 \text{ and } j \leq N\,.
 \end{cases}
 \label{eq:covariance_matrix}
\end{equation}
We now write the invert of this matrix $M=C^{-1}$ and $M_s$ the
top-left most $N\times N$ sub-matrix of $M$. In other words, this
sub-matrix corresponds to the first $N$ lines and $N$ columns of $M$.
The probability of the vector $(\mathbf{\widehat{U}},V)$ is thus:
\begin{equation}
 P(\mathbf{\widehat{U}},\widehat{V}) = \frac{1}{(2\pi)^{(N+1)/2} \sqrt{\det C}}
 \exp\left(-\frac{1}{2} {}^{t}\mathbf{\widehat{U}} M_s \mathbf{\widehat{U}} -
  \widehat{V} M_{i,(N+1)} \widehat{U}_i - \frac{1}{2} M_{(N+1),(N+1)} \widehat{V}^2 \right)\,.
\end{equation}
Now we use Bayes theorem to compute $P(\mathcal{U}|V)$:
\begin{equation}
 P(\mathcal{U}|\mathcal{V}) = \frac{P(\mathcal{U},\mathcal{V})}{\int P(\mathcal{U},\mathcal{V})\text{d}\mathcal{U}}\,.
\end{equation}
Before computing the integral in the denominator, we need to rewrite
the argument of the exponential:
\begin{equation}
 {}^{t}\mathbf{\widehat{U}} M_s \mathbf{\widehat{U}} + 2 \widehat{V} M_{i,(N+1)} \widehat{U}_i +
 M_{(N+1),(N+1)} \widehat{V}^2 = {}^{t}\mathbf{\widehat{W}} M_s \mathbf{\widehat{W}} + \alpha \widehat{V}^2\,. \label{eq:exp_argument}
\end{equation}
with
\begin{eqnarray}
 \mathbf{\widehat{W}} & = &\mathbf{\widehat{U}} + \mathbf{\widehat{A}}\,, \\
 \widehat{A}_i & = & (M_s^{-1})_{i,j} M_{N+1,i} \widehat{V} = -C_{N+1,i} \widehat{V} = -\Gamma_i \widehat{V}\,,\\
 \alpha & = & M_{N+1,N+1} - M_{N+1,i} M_{s,i,j} M_{N+1,j}\,.
\end{eqnarray}
We may now compute the denominator:
\begin{equation}
 \int_{\mathbf{\widehat{U}}} P(\mathbf{\widehat{U}},\widehat{V}) \text{d}\mathbf{\widehat{U}} =
 \int_{\mathbf{\widehat{W}}} \frac{\sqrt{\det M}}{(2\pi)^{(N+1)/2}} \exp\left( -\frac{1}{2} {}^{t}\mathbf{\widehat{W}} M_s
 \mathbf{\widehat{W}} - \frac{1}{2} \alpha \widehat{V}^2\right) = \frac{\sqrt{\det
     M}}{\sqrt{\det M_s}} \exp\left(-\frac{1}{2} \alpha \widehat{V}^2\right)\,.
\end{equation}
We may rewrite the argument of the exponential in the numerator using \eqref{eq:exp_argument} and deduce the conditional probability 
\begin{equation}
 P(\mathbf{\widehat{U}}|\mathbf{\widehat{V}}) = \frac{\sqrt{\det M_s}}{(2\pi)^{N/2}}
 \exp\left(-\frac{1}{2} ({\widehat{U}}_i - \Gamma_i \widehat{V})(\widehat{U}_j - 
 \Gamma_j \widehat{V}) M_{s,i,j}\right)\,.
\end{equation}
This conditional probability is written in terms of the normalized
variables $\mathbf{\widehat{U}}$ and $\widehat{V}$.  The probability
may be rewritten in terms of the original random variables by an appropriate change of
variables:
\begin{equation}
 P({\bf U}|V) = \frac{\sqrt{\det M_s}}{(2\pi)^{N/2}} \left(\prod_{i=1}^N \frac{1}{\sigma_i}\right)
 \exp\left( -\frac{1}{2} (U_i/\sigma_i - \Gamma_i V/\sigma_V)(U_j/\sigma_j - 
 \Gamma_j V/\sigma_V) M_{s,i,j} \right)\,. \label{eq:conditional_gaussian}
\end{equation}

\section{Appendix B\\[.5cm] Comparison of linear theory and MAK reconstruction of dipole}
\label{app:lin_vs_mak}

We have demonstrated, in contrast to previous reconstruction of 2MRS peculiar velocities \citep{Erdogdu2005,Erdogdu2006}, that our method  agrees well with recent works which make direct use of observed peculiar velocities \citep{Feldman08}. 
In this appendix, the difference between linear theory and MAK determination of bulk flow is studied.

There are mainly two ways of implementing linear theory for getting the velocity of the Local Group. One method is the flux-weighed method \citep[as recently done for 2MRS in][]{Erdogdu2005}. The second method uses linearized fluid equation to infer the velocity field from the density field. The advantage of the former method is that it bypasses the redshift space distortions, which could be problematic for objects highly extended in redshift space (e.g. Virgo cluster). The disadvantage of this method is that it is limited to the velocity of the Local Group. It does not yield the peculiar velocity field and especially in the local neighborhood. In addition, to study the impact of farther and farther structures on the Local Group, one often substitutes distances by redshifts. 

The second way of obtaining peculiar velocity field from linear theory relies on linearized fluid equations. The disadvantage of this method is that the Local Group velocity becomes sensitive to redshift space distortions. The advantage is that we obtain the 3d velocity field everywhere and not just at our location. In this appendix, we compare this way of implementing linear theory with MAK. 

To estimate the velocity of the Local Group with linear theory, we follow the same procedure for accounting for redshift distortions as for MAK, except that we use the linear relation between peculiar velocity field and the density field given by:
\begin{equation}
  \text{div}\,{\bf v} = -\beta H \delta,
\end{equation}
where $\beta = \Omega_\text{m}^{5/9} / b$, $b$ is the bias, $H$ the Hubble constant and $\delta$ the density fluctuations. We use a Gaussian smoothing filter of 2.5\Mpch{} of radius and increase progressively $\beta$ to increase numerical stability \citep[e.g][]{Pike05}.

To study the difference between linear theory and MAK, we look at four most prominent structures (listed in Table~\ref{tab:structures}) in 2MRS catalog and study their effect on the Local Group motion.  We evaluate the Local Group velocity in the presence and in the absence of these four structures using both linear theory and MAK.
We computed the velocity of the
Local Group based on a 60\Mpch{} cut of the 2MASS Redshift Survey and in
the rest frame of this volume. 

\begin{figure*}
  \begin{center}
    \includegraphics[width=.8\hsize]{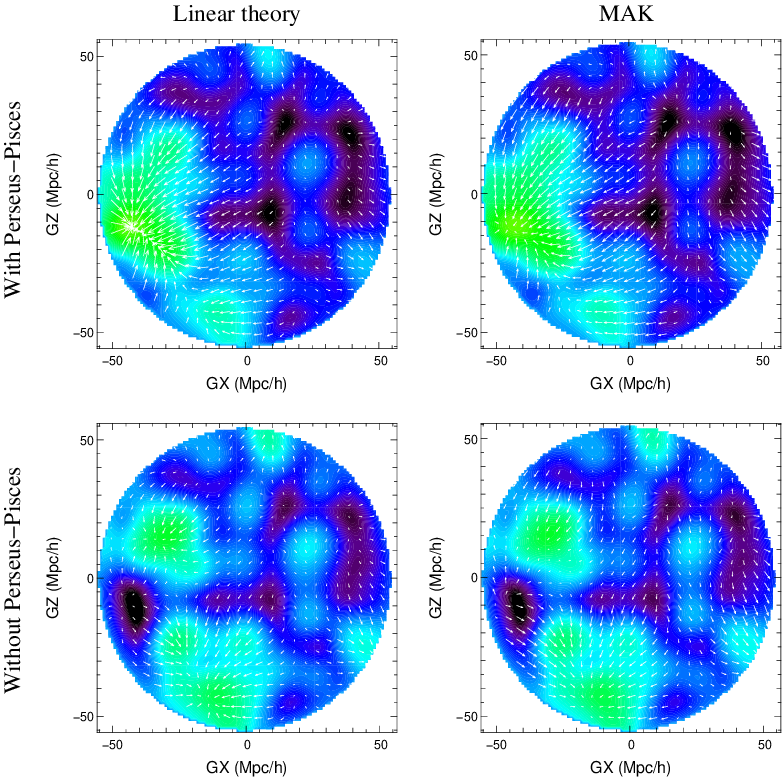}
  \end{center}
  \caption{\label{fig:pp_slice} {\it Reconstructed velocity field in the region of the Perseus-Pisces in Galactic coordinates} -- We give here the reconstructed velocity field using either linear theory (left panels) or by solving the Monge-Amp\`ere-Kantorovitch problem (right panels) for two cases: with the Perseus-Pisces supercluster (top row), and removing the galaxies of the Perseus-Pisces supercluster (bottom row). In all cases, we limited the reconstruction to a volume of radius 60\Mpch{} from the observers. The slice is centered on the Perseus cluster and does not contain Milky Way. The slice corresponds  actually to GY=25\Mpch{}.   }
\end{figure*}

Table~\ref{tab:structure_impact} demonstrates the effect of these four structures on the Local Group velocity. We note that linear theory overestimates the Local Group motion due to lower mass nearby objects and underestimates the effect of high mass farther objects. This deserves a full theoretical understanding in the framework of perturbative theory which is beyond the scope of this work. Since Perseus-Pisces is the most massive object in our study, we also show in Figure~\ref{fig:pp_slice} the effect of this structure on the entire velocity field, as reconstructed using linear theory and MAK. The large scale flows are  similar in the two methods and, as expected, the small scale flows are different.

\begin{table}
  \caption{\label{tab:structures} Structures probed for comparing with linear theory}
  \begin{center}
    \begin{tabular}{cccccc}
      \hline
      Structure & $z_\text{min}$ & $z_\text{max}$ & $l$ & $b$ & Opening angle \\
                & (\kms) & (\kms) & (deg) & (deg) & (deg) \\
      \hline
      \hline
      Virgo & no minimum & 3000 & 279 & 74 & 20 \\
      Hydra & 2000 & 4300 & 269 & 26 & 15 \\
      Centaurus & 2000 & 4300 & 302 & 21 & 15 \\
      Perseus-Pisces & 4000 & 6000 & 150 & -13 & 20 \\
      Fornax & no minimum & 1600 & 240 & -50 & 20 \\
      \hline
    \end{tabular}
  \end{center}
\end{table}

\begin{table}
  \caption{\label{tab:structure_impact} Impact of structures on the final velocity of the Local Group, smoothed with a Gaussian kernel of 5\Mpch{} radius}
  \begin{center}
    \begin{tabular}{cccccc}
      \hline
             &           & & \multicolumn{3}{c}{$\delta {\bf v}_\text{S,method}$} \\
      Method & Structure (S) & Mass & $|\delta V|$ & $l$ & $b$ \\
             &           & ($\Msun$) & (\kms)  & (deg) & (deg)  \\
      \hhline{*{6}=}
      MAK    & Virgo     & $1.3\times 10^{15}$ & 89      & 302 & 88  \\
      Linear &           & & 230     & 297 & 68 \\
     \hhline{*{6}-}
      MAK    & Fornax    & $3.6\times 10^{14}$ &52      & 261 & -15 \\
      Linear &           & &94      & 271 & -1  \\
     \hhline{*{6}-}
      MAK    & Hydra-Centaurus & $4.3\times 10^{15}$ & 88      & 281 & 35 \\
      Linear &           & & 189     & 297 & 39 \\
      \hhline{*{6}-}
      MAK    & Perseus-Pisces & $8.9\times 10^{15}$ & 105     & 167 & 10  \\
      Linear &           & & 52      & 139 & 48 \\
      \hhline{*{6}-}
    \end{tabular}
  \end{center}
  {\it Note -- } We are giving in this table the direction and amplitude of the velocity vector corresponding
  to the velocity of the Local Group when all structures are present minus its velocity when we removed
  the indicated structure. The structures are defined in Table~\ref{tab:structures}. The comparison is
  done for both the result given by MAK and the linear theory.
\end{table}

\begin{figure}
  \begin{center}
    \includegraphics[width=.8\hsize]{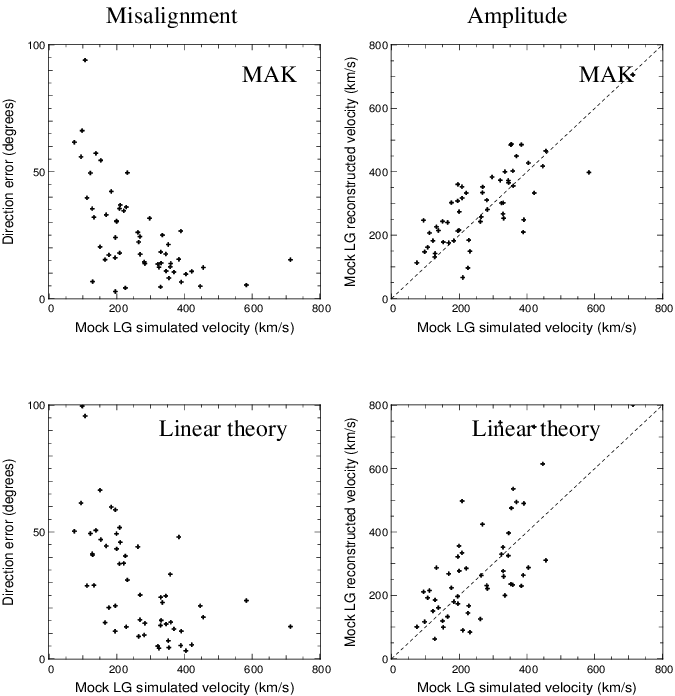}
  \end{center}
  \caption{ \label{fig:mockobservers} {\it Comparing MAK to linear theory through Monte Carlo tests: reconstructed velocity vs. simulated velocity} -- We represent here the comparisons between reconstructed velocities (top row MAK and bottom row linear theory) and simulated velocities of 57 mock-observers in 57 mock catalogs. The velocity fields (simulated, reconstructed by MAK, reconstructed by linear theory) were smoothed to 2.5\Mpch{} with a Gaussian kernel. Left panels: direction misalignment vs amplitude of the simulated velocity. Right panels: amplitude of the reconstructed velocity vs amplitude of the simulated velocity. The mock-observer velocities reconstructed using MAK (top row) show $\sim$30\% less dispersion with respect to linear theory.  }
\end{figure}

We have done a more extensive test of the difference between linear theory and MAK reconstruction through a Monte-Carlo test. We have taken fifty observers, randomly located in the simulation, and built a mock catalog centered on each of these observers. These mock catalogs are extracted from the $\Lambda$CDM simulation described in \cite{lavaux08}. Each mock catalog is limited in depth to 60\Mpch{} and mock galaxies and groups are put in redshift space. We calculated the velocity of the mock observer given by the linear theory and by MAK and then compared these to the true value given by the simulation. The result is shown in Figure~\ref{fig:mockobservers}. MAK presents a dispersion $\sim$30\% less than linear theory. (For other comparisons with linear theory see Lavaux 2009.)

\clearpage
\section{Appendix C\\[.5cm] Adaptive smoothed interpolation}
\label{app:weigh}

We make use of a technique inspired from adaptive smoothing \citep{VelGrav07},
while changing the weight such that we make an smoothed interpolation of the
required field (in this work the velocity field). We put $A({\bf y})$ the
underlying continuous field, and $A_i$ its value at the position ${\bf
  x}_i$. We now define the smoothed interpolated field 
\begin{equation}
  \tilde{A}({\bf y}) = \frac{1}{N({\bf y})} \sum_{i=1}^M
  W\left(\frac{{\bf y}-{\bf x}_i}{R({\bf y})}\right) A_i \label{eq:smooth_interp}
\end{equation}
with $M$ the number of tracers on which the smoothed interpolation is done,
\begin{equation}
  N({\bf y}) = \sum_{i=1}^M W({\bf y}-{\bf x}_i)
\end{equation}
the normalization coefficient of the smoothing and
\begin{equation}
  W(x) = \left\{ \begin{array}{ll}
                 1 - \frac{3}{2} x^2 + \frac{3}{4} x^3 & \;\; x \leq 1 \\
                 \frac{1}{4} (2-x)^2 & \;\; 1 < x \leq 2 \\
                 0 & \;\;\text{otherwise}
  \end{array}\right.
\end{equation}
the smoothing kernel used. In practice, in the above smoothing procedure, we
set $M=32$, and $R({\bf y})$ is half the distance to the 32-th neighbors from
${\bf y}$.

It can be shown that for class 1 function\footnote{Class 1 functions are
  continuous and their derivative is also continuous.} and an unbiased distribution of
tracers of $A({\bf y})$ this smoothing procedure converges
normally on any volume $V$. Locally, the sum \eqref{eq:smooth_interp} may be
expanded in
\begin{eqnarray}
  \tilde{A}({\bf y}) & = & \frac{1}{N({\bf y})} \sum_{i=1}^M W\left(\frac{{\bf y}-{\bf
      x}_i}{R({\bf y})}\right)  \left( A({\bf y}) + ({\bf \nabla} A)({\bf y}) \cdot ({\bf
    x}_i-{\bf y}) \right) \\
   & = & A({\bf y}) + ({\bf \nabla} A)({\bf y})\cdot \left(\sum_{i=1}^M
  \frac{1}{N({\bf y})} W\left(\frac{{\bf y}-{\bf
      x}_i}{R({\bf y})}\right) ({\bf x}_i-{\bf y}) \right)\,.
\end{eqnarray}
The second term on the right-hand-side may be bound by:
\begin{equation}
   ({\bf \nabla} A)({\bf y})\cdot \left(\sum_{i=1}^M
  \frac{1}{N({\bf y})} W\left(\frac{{\bf y}-{\bf
      x}_i}{R({\bf y})}\right) ({\bf x}_i-{\bf y}) \right) \leq 2 ||{\bf \nabla}
  A||_{+\infty,V} M R\,.
\end{equation}
As $R \propto n^{-1/3}$ with $n$ the number density of tracers and that
$||{\bf \nabla} A||_{+\infty}$ is bounded on any volume $V$, the above quantity
decreases to zero when $n\rightarrow +\infty$ and the interpolation
$\tilde{A}$ converges to $A$. 

\section{Appendix D\\[.5cm] Test of 2MRS reconstructed velocities with observed distances within 3,000\kms{}}
\label{app:loc_bulk}

\begin{figure}[t]
  \includegraphics*[width=\hsize]{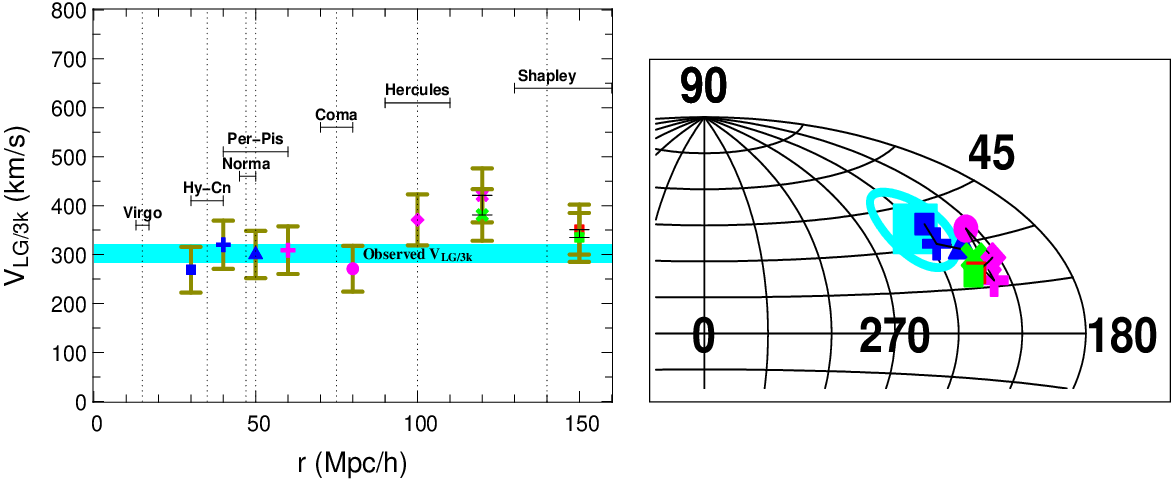}
  \caption{{\it The origin of motion of the Local Group in the rest frame of galaxies
      in the 3,000~\kms (3k) volume,$V_{\rm LG/3k}$}: The observed
    amplitude and direction of the $V_{\rm LG/3k}$ are shown by the
    horizontal cyan band in the lower panel and the solid cyan large
    square in the top panel, respectively. The observational error on
    the direction is shown on the top panel by the thick cyan
    ellipse. Each reconstructed point is affected by a $22^\circ$
    error. The reconstructions indicate that most of the motion is
    generated by the structures within this volume, in agreement with
    the observations.  It is seen that most of the motion is generated
    within 3,000~\kms{}. The jump at around 120\Mpch{} is most
    probably caused by the severe incompleteness of 2MRS at this
    distance. The represented error bar have been computed by adding
    in quadrature the 9\% systematic error bar due to the uncertainty
    on $\Omega_\text{m}$ and a random reconstruction error of 70~\kms. Incompleteness
    results in uncertainties illustrated by the black error bars on
    data points beyond 120\Mpch{} in the bottom panel.
    \label{fig:dipole3k}}
\end{figure}

The reconstructed velocities of objects lying only within the
3,000~\kms{} radius can be compared to the measured distances given by
the 3k distance catalog \citep{TullyVoid07}. The measured distances give a velocity of the
Local Group with respect to the 3,000~\kms{} volume of $V_\text{LG/3k} = 302\pm
22$~\kms{}, $l = 241\pm 7$, $b = 37 \pm 6$.
The observation indicates that most of this velocity comes from
the push from the Local Void and the gravitational pull of the Virgo
cluster \citep{TullyVoid07}. The velocity of the Local Group with respect to the 3,000~\kms{}
volume ($V_{\rm LG/3k}$) is obtained using our reconstructed 2MRS
velocities. The reconstructed $V_{\rm LG/3k}$ is compared with the
observed value in Fig.~\ref{fig:dipole3k}. The coordinates of the
reconstructed dipoles are given in
Table~\ref{table:dipole3k}. 
The influence of other structures, like the
  Hydra-Centaurus-Norma at 40-50\Mpch, seems marginal in this rest frame.
So we conclude that
the reconstruction indeed shows that the $V_{\rm LG/3k}$ motion seems mainly
generated within the 3,000~\kms{} volume.

\begin{table}
  \caption{\label{table:dipole3k} Comparison between the observed and the reconstructed velocity of the Local Group in the 3,000~\kms{} rest frame}
  \begin{center}
    \begin{tabular*}{.85\hsize}{@{\extracolsep{\fill}}cccccccc}
      \multicolumn{8}{c}{
        \begin{tabular}{lp{.5\hsize}}
          Observed velocity: & $V_\text{LG/3k}=302$~\kms, $l=241$, $b = 37$\\
        \end{tabular}
      }\\
      \multicolumn{8}{c}{
        $v_{\text{LG/3k},x} = -125 \pm 24$~\kms{}, $v_{\text{LG/3k},y} = -210 \pm 21$~\kms{}, $v_{\text{LG/3k},z} = 181 \pm 18$~\kms{}
      } \\
      \\
      \hline
      \hline
      \multirow{2}{1.5cm}{\hfill $R_\text{rec}$ \hfill}  & \multicolumn{7}{c}{$V_\text{LG/3k}$} \\
      \cline{2-8}
       & \hspace{\myextralen}$v_x$\hspace{\myextralen} & \hspace{\myextralen}$v_y$\hspace{\myextralen} & \hspace{\myextralen}$v_z$\hspace{\myextralen} & $|V|$ & $l$ & $b$ & angular \\
      (\Mpch) & (\kms) & (\kms) & (\kms) & (\kms) & (deg) & (deg) & separation \\
      \hline
      30 & $-122 \pm 71 $   & $-173 \pm 71 $ & $166 \pm 71 $ & 269 & 234 & 38 & 5\\
      40 & $-150 \pm 71$   & $-228 \pm 73$ & $166 \pm 71$ & 320 & 236 & 31 & 7 \\
      50 & $-183 \pm 72$  & $-187 \pm 72$ & $145 \pm 71$ & 300 & 225 & 28 & 16 \\
      60 & $-232 \pm 73$  & $-182 \pm 72$ & $90 \pm 70$   & 309 & 218 & 17 & 28\\
      80 & $-183 \pm 72$  & $-128 \pm 71$ & $154 \pm 71$ & 271 & 215 & 34 & 21\\
      100 & $-282 \pm 73$ & $-176 \pm 72$ & $152 \pm 71$ & 371 & 213 & 24 & 27 \\
      120 & $-311 \pm 73$ & $-230 \pm 73$ & $164 \pm 71$ & 421 & 216 & 23 & 25\\
      150 & $-231 \pm 73$ & $-234 \pm 73$ & $124 \pm 71$ & 351 & 225 & 20 & 21\\
      \hline
      \hline 
    \end{tabular*}
  \end{center}
\end{table}

\end{document}